\documentclass[preprint,11pt]{elsarticle}

\usepackage{amsmath}
\usepackage{array}
\usepackage{amssymb}
\usepackage{graphicx,epsfig,psfrag}
\usepackage{fancyhdr}
\usepackage{multirow}
\usepackage{mathrsfs}
\usepackage{tabu}
\usepackage{stmaryrd}
\usepackage{caption}
\usepackage{xfrac}

\newcommand\bs[1]{\boldsymbol{#1}}

\usepackage{graphicx}
\usepackage{amssymb}

\usepackage{lineno}

\newlength{\figwidthfull}
\newlength{\figwidthhalf}
\newlength{\figwidththird}
\newlength{\figwidthsmall}
\newlength{\figwidthsmallsmall}
\newlength{\figspace}

\usepackage[colorlinks=true,linkcolor=black, citecolor=blue, urlcolor=blue]{hyperref} 
\usepackage{subfigure}

\usepackage{algorithm,algcompatible,amsmath}
\algnewcommand\INPUT{\item[\textbf{Input:}]}%
\algnewcommand\OUTPUT{\item[\textbf{Output:}]}%
\algnewcommand{\Initialize}[1]{%
  \State \textbf{Initialize:}
  \Statex \hspace*{\algorithmicindent}\parbox[t]{.8\linewidth}{\raggedright #1}
}

\algnewcommand\algorithmicswitch{\textbf{switch}}
\algnewcommand\algorithmiccase{\textbf{case}}
\algnewcommand\algorithmicassert{\texttt{assert}}
\algnewcommand\Assert[1]{\State \algorithmicassert(#1)}%

\algdef{SE}[SWITCH]{Switch}{EndSwitch}[1]{\algorithmicswitch\ #1\ \algorithmicdo}{\algorithmicend\ \algorithmicswitch}%
\algdef{SE}[CASE]{Case}{EndCase}[1]{\algorithmiccase\ #1}{\algorithmicend\ \algorithmiccase}%
\algtext*{EndSwitch}%
\algtext*{EndCase}%
\begin{document}
\captionsetup[figure]{labelfont={bf},name={Fig.},labelsep=period}
\begin{frontmatter}
\title{Adaptive and Parallel Multiscale Framework for Modeling Cohesive Failure in Engineering Scale Systems}
\author[add1]{Sion Kim}
\author[add2]{Ezra Kissel}
\author[add1]{Karel Matou\v{s} \corref{cor1}}
\address[add1]{Department of Aerospace and Mechanical Engineering, University of Notre Dame, IN, 46556}
\address[add2]{Lawrence Berkeley National Laboratory, Energy Sciences Network, Berkeley, CA, 94720}
\cortext[cor1]{Corresponding author. E-mail address: \href{mailto:kmatous@nd.edu}{kmatous@nd.edu} (K. Matou\v{s})}
\begin{abstract}
The high computational demands of multiscale modeling necessitate
advanced parallel and adaptive strategies. To address this challenge,
we introduce an adaptive method that utilizes two microscale models
based on an offline database for multiscale modeling of curved
interfaces (e.g., adhesive layers). This database employs nonlinear
classifiers, developed using Support Vector Machines from microscale
sampling data, as a preprocessing step for multiscale
simulations. Next, we develop a new parallel network library that
enables seamless model selection with customized communication layers,
ensuring scalability in parallel computing environments. The
correctness and effectiveness of the hierarchically parallel solver
are verified on a crack propagation problem within the curved adhesive
layer. Finally, we predict the ultimate bending moment and adhesive
layer failure of a wind turbine blade and validate the solver on a
difficult large-scale engineering problem.
\end{abstract}
\end{frontmatter}
\section{Introduction}
The rapid expansion of composite materials in aerospace, wind energy,
automotive, and other industries has stimulated significant research
interest in multiscale methods. In general, multiscale methods
have the ability to construct the constitutive response from data at
lower scales and can efficiently transform epistemic uncertainty
(i.e., reducible uncertainty) in the parameters of coarse-grained
models into the aleatoric uncertainty (i.e., irreducible uncertainty)
of finer-grained models. Review articles by Matou\v{s} et
al.~\cite{JCP}, Fish et al.~\cite{Fish2021}, Schlick et
al.~\cite{Schlick_2021}, and Geers et al.~\cite{geers2010multi}
highlight important milestones in multiscale modeling and
delineate outstanding challenges.

Multiscale methods, and homogenization in particular, trace back
to micromechanics. Early models, such as those by Sachs~\cite{sachs1928},
Taylor~\cite{taylor1938}, Eshelby~\cite{eshelby1957},
Willis~\cite{willis1977}, and Hill~\cite{hill1972constitutive}, among
others, were developed to estimate effective macroscopic constitutive
equations. The ``homogenization'' method was originally conceived by
Ivo Babu\v{s}ka~\cite{babuska1977} and many following works provided rigor to its foundation (see the work of Benssousan et
al.~\cite{benssousan1978}, and
Sanchez-Palencia~\cite{sanchezpalencia1980}, for example).

Since its early development, many powerful homogenization methods have
been developed for a variety of engineering problems, with work by
Fish~\cite{fish2009}, Geers et al.~\cite{Geers2016}, Ghosh et
al.,~\cite{ghosh1996}, Khisaeva and
Ostoja-Starzewski~\cite{khisaeva2006}, Temizer and
Zohdi~\cite{temizer2007b}, Terada et al.~\cite{terada2000simulation},
Miehe et al.~\cite{miehe1999computational}, Ponte
Casta\~neda~\cite{pontecastaneda2002}, Michel et
al.~\cite{michel1999}, Feyel and Chaboche~\cite{feyel2000fe2},
Matou{\v{s}} et al.~\cite{matouvs2008multiscale}, and many other
scholars. Today, homogenization is regarded as a well-developed
discipline with an impact on industry and academia alike.

\begin{figure}[!htb]
\centering
\includegraphics[width=0.9\textwidth]{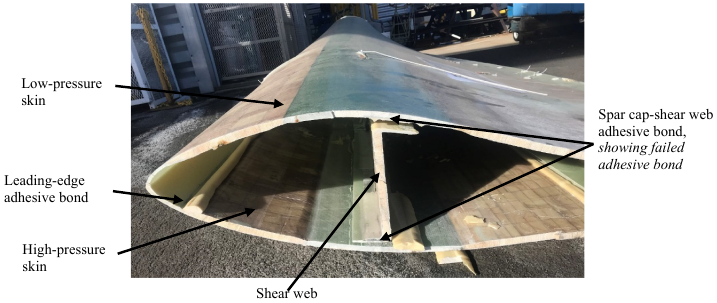}
\caption{National Rotor Testbed 13 m wind turbine blade. (from Murray et al.~\cite{murray2019fusion})}\label{figure:blade}
\end{figure}
In this article, we are particularly interested in the multiscale
modeling of adhesive layers. Adhesive layers are critical in many
engineering applications. Fig.~\ref{figure:blade} shows progressive
debonding of a 13 m wind turbine blade designed by Sandia National
Laboratories~\cite{kelley2015aerodynamic} and structurally tested at
the National Wind Technology Center~\cite{gage2021laboratory}. During
the ultimate static test, the adhesive joint failed between the shear
web and the spar cap on the low-pressure side. This highlights the
modeling challenges of capturing the structural response of the blade
as well as the highly nonlinear response of the adhesive. To address
the modeling challenges of adhesive failure, Matou\v{s} et
al.~\cite{matouvs2008multiscale,Kulkarni3,kulkarni2009multi} developed
a multiscale scheme for interfaces, which was later expanded to finite
strains~\cite{hirschberger2009computational,MOSBY2015-1}. The first
three-dimensional (3D) hierarchically parallel implementation was
developed by Mosby and Matou\v{s}~\cite{MOSBY2014,MOSBY2015-2} and
successfully applied to large multiscale problems. 

Unfortunately, the computational complexity of multiscale modeling remains a bottleneck, and many researchers have focused on developing
Reduced-Order Models (ROM). Oskay and Fish~\cite{oskay2007eigendeformation}
developed eigendeformation-based reduced order homogenization. Yvonnet
and He developed a reduced model multiscale
method~\cite{Yvonnet2007}. Michel and Suquet developed a nonuniform
transformation field analysis~\cite{michel2003}. van Tuijl et
al.~\cite{Tuijl_2017} developed a wavelet based reduced order
model. Bhattacharjee and
Matou\v{s}~\cite{bhattacharjee2016nonlinear,BHATTACHARJEE2020112657}
developed a manifold-based reduced order model. Le et al.~\cite{NME:NME4953}
used neural networks for computational homogenization. Beel and Fish
proposed solver-free reduced order
homogenization~\cite{BEEL2024116932}. Adaptive multiscale modeling is
also
popular~\cite{fan2010adaptive,ghosh2001,budarapu2014adaptive,chung2016adaptive,bauman2009adaptive,ODEN1997367,oden2000estimation,vemaganti2001estimation}. Furthermore,
we highlight the impact of machine learning on multiscale
modeling~\cite{peng2020multiscale,WANG2018337,chan2018machine,Bishara_2023,LOGARZO2021113482}.

In this work, we develop a novel adaptive and parallel multiscale
framework that combines both model reduction and efficient parallel
implementation. In particular, we focus on Computational
Homogenization (CH) and microscale model
selection~\cite{bartlett2002model}, which are crucial for solving real
industrially relevant problems, like in Fig.~\ref{figure:blade}. For large
engineering problems, multiscale simulations are usually needed in
only a small portion of the computational domain (i.e., the
damage-prone regions of the adhesive layer), and the rest of the structure can be
modeled using a ROM or a phenomenological constitutive model. However, the matter of how to
effectively select from a list of models is a nontrivial task, especially in a
parallel multiscale environment where each model requires different
computational resources. To address this issue, we rely on Support
Vector Machines (SVMs)~\cite{cortes1995support,drucker1996support},
which are supervised max-margin models with associated learning algorithms
that analyze data for regression analysis and classification. SVMs
offer potential for enhancing statistical analysis in multiscale
modeling due to their robust mathematical foundations and
scalability~\cite{makridakis2018statistical,hearst1998support,steinwart2008support}. We
extend the 3D multiscale modeling of interfaces~\cite{MOSBY2014} to
problems with curvature, which requires a careful computational
treatment of the unit cell normal. Moreover, we develop a novel
multiscale network library that allows uninterrupted dynamic
microscale model selection and computer resource allocation in a
parallel setting. We verify the framework on the Double-Cantilever
Beam (DCB) with a curved interface and validate it on a difficult National
Rotor Testbed (NRT) engineering problem.

The remainder of this paper is organized as follows. In
Section~\ref{sec:2}, we overview the governing equations for CH and
SVMs. Section~\ref{sec:3} explores the adaptive strategy for selecting
microscale models. In
Section~\ref{sec:4}, we introduce the supporting network for parallel and
adaptive multiscale simulation. Following this, Section~\ref{sec:5} discusses the
constitutive model alongside the numerical
implementation. Section~\ref{sec:6} presents numerical
examples. Finally, the conclusions are drawn in Section~\ref{sec:7}.
\section{Governing equations}\label{sec:2}
In this section, we present computational homogenization for
interfaces and SVMs for model selection. Building upon the previous
multiscale work outlined in~\cite{MOSBY2015-2}, we develop and
incorporate a novel adaptive strategy suitable for high-performance
computing (i.e, parallel computing). The following subsections will
cover both macro and micro scale problems, with an overview of Support
Vector Regression (SVR), which is a variant of the SVM, as the
methodology for microscale model adaption.

\subsection{Computational homogenization for interfaces}
\begin{figure}[!htb]
\centering
\includegraphics[width=0.8\textwidth]{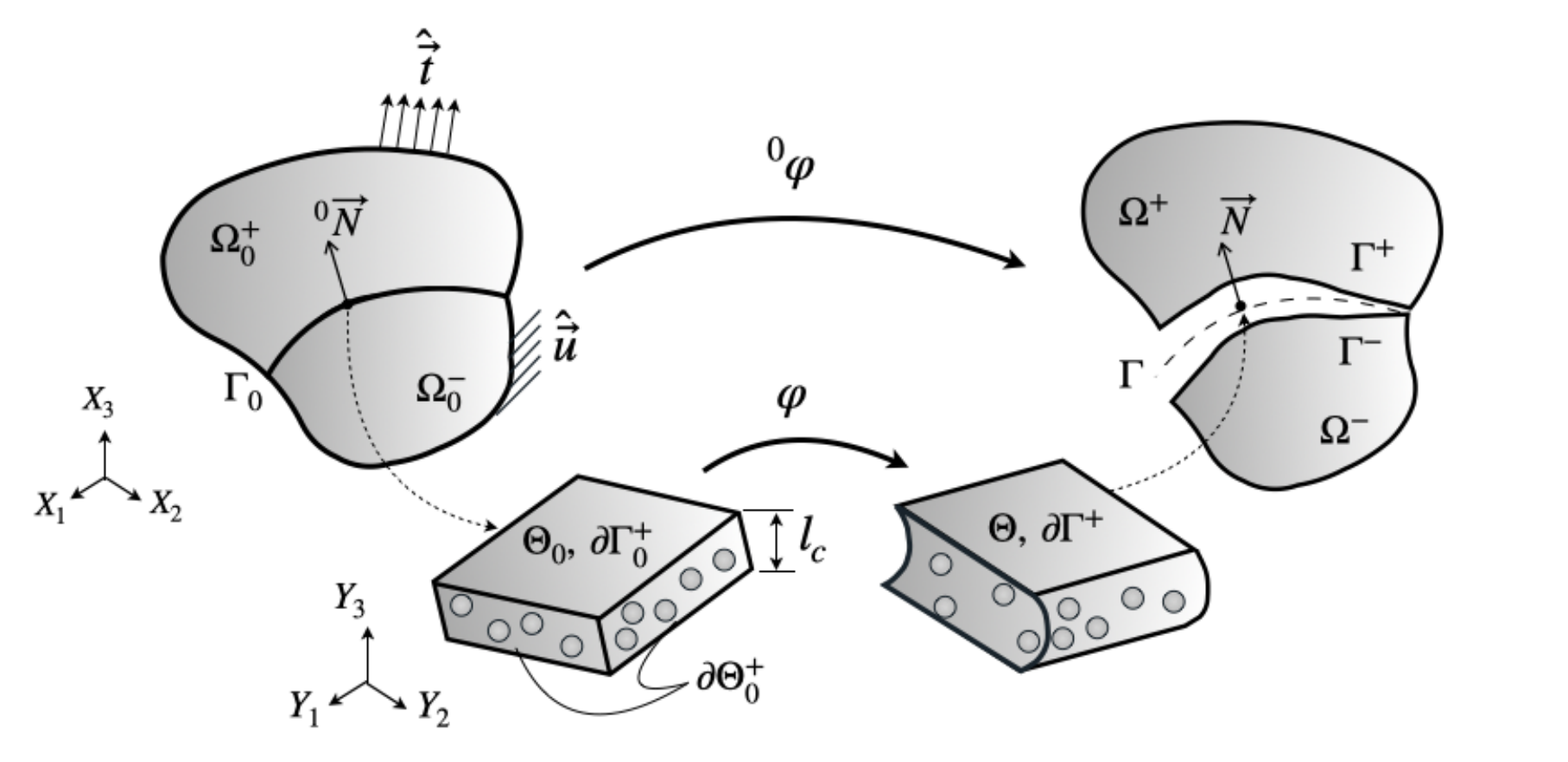}
\caption{Kinematics of multiscale cohesive modeling in a finite strain
setting. $\Omega_0^\pm$ are the macroscale domains, while $\Theta_0$ represents the microscale domain.} 
\label{figure:potato_interface}
\end{figure}

For clarity of presentation, we revisit multiscale cohesive
modeling in a finite strain setting as originally developed
in~\cite{matouvs2008multiscale,MOSBY2014,MOSBY2015-1}. The
position of a macroscopic material point is defined as
$\vec{X}\in\Omega_0$ in a body $\Omega_0\subset\mathbb{R}^3$, and the
position of a microscopic material point is defined as
$\vec{Y}\in\Theta_0$ in a microstructure
$\Theta_0\subset\mathbb{R}^3$, as shown in
Fig.~\ref{figure:potato_interface}. The adhesive layer is represented
by the domain $\Gamma_0\subset\mathbb{R}^2$ with a positive thickness
$l_c$ linking two adherends $\Omega_0^\pm$ and a normal to the
interface ${}^0\vec{N}$. The boundary is decomposed into $\partial
\Omega^u_0$ and $\partial \Omega^{\text{t}}_0$, for the applied
displacement $\hat{\vec{u}}$ and traction $\hat{\vec{t}}$ vectors
respectively, satisfying the relations $\partial
\Omega_0=\partial\Omega_0^u \cup \partial\Omega_0^{\text{t}}$ and
$\partial \Omega_0^u\cap \partial\Omega_0^{\text{t}}=\emptyset$.

Within the multiscale cohesive modeling framework, the average deformation
gradient on the interface $\Gamma_0$ is defined by the jump across the
layer $\llbracket{}^0\varphi\,
\rrbracket={}^0\varphi^+-{}^0\varphi^-=\llbracket{}^0\vec{u}\,\rrbracket$. Thus,
the macroscale deformation gradients read
${}^0\bs{F}=\nabla_{\vec{X}}{}^0\varphi=\bs{1}+\nabla_{\vec{X}}{}^0\vec{u}$
for the macroscale domain $\Omega_0^\pm$, and
${}^0\bs{F}=\bs{1}+\sfrac{1}{l_c}\llbracket{}^0\vec{u}\,\rrbracket\otimes{}^0\vec{N}$
for the interface $\Gamma_0$. The macroscale boundary value problem
stems from the linear momentum balance. By neglecting inertial
terms and body forces, we obtain 
\begin{align}
    \nabla_{\vec{X}}\cdot({}^0\boldsymbol{F}\,{}^0\boldsymbol{S})=\vec{0}\quad&\in \Omega_0, \label{eq:pde_to_solve_0} \\
    {}^0\bs{P}\cdot{}^0\vec{N}=\hat{\vec{t}}\quad&\text{on  } \partial \Omega_0^{\text{t}},\\
    {}^0\vec{u}=\hat{\vec{u}}\quad&\text{on  }\partial \Omega_0^u,\\
    {}^0\vec{t}^++{}^0\vec{t}^-=\vec{0}\quad&\in \Gamma_0,
\end{align}
where ${}^0\bs{P}={}^0\bs{F}\,{}^0\bs{S}$ and
${}^0\bs{S}=2\partial\,{}^0W({}^0\bs{C})/\partial\,{}^0\bs{C}$ are the
first (PK1) and the second Piola-Kirchhoff (PK2)
stress tensors, respectively. Here, PK2 is derived from the macroscale
strain energy density function ${}^0W({}^0\bs{C})$ and
${}^0\bs{C}={}^0\bs{F}^{\top}\,{}^0\bs{F}$ denotes the macroscopic
right Cauchy–Green deformation tensor.

The microscale deformation gradient is represented as
$\boldsymbol{F}={}^0\boldsymbol{F}+\nabla_{\vec{Y}}{}^1\vec{u}$. By
neglecting the body forces and assuming no prescribed traction, we
obtain the microscale boundary value problem
\begin{align}
    \nabla_{\vec{Y}}\cdot(\boldsymbol{F}\,{}^1\boldsymbol{S})=\vec{0}\quad&\forall\,\vec{Y}\in \Theta_0,\label{eq:pde_to_solve}\\
    \varphi=\hat{\varphi}\quad&\text{on  } \partial \Theta_0,
\end{align} 
where ${}^1\bs{S}=2\partial {}^1W(\bs{C})/\partial \bs{C}$ is the
microscale PK2 stress tensor, ${}^1W(\bs{C})$ represents the microscale strain
energy density function, and $\bs{C}=\bs{F}^{\top}\bs{F}$ is the
microscale right Cauchy-Green deformation tensor.

We bridge the macroscale and microscale boundary value problems
through the Hill-Mandel energy equivalence~\cite{hill1972constitutive}
\begin{align}
    \inf_{\llbracket{}^0\vec{u}\,\rrbracket}{}^0W(\llbracket{}^0\vec{u}\,\rrbracket)=\inf_{\llbracket{}^0\vec{u}\,\rrbracket}\inf_{{}^1\vec{u}}\frac{l_c}{|\Theta_0|}\int_{\Theta_0}{}^1W({}^0\bs{F}+\nabla_{\vec{Y}}{}^1\vec{u})\,\text{d}\Theta_0,\label{eq:hills}
\end{align}
which is usually formulated for dissipative processes in terms of work conjugated pairs~\cite{Geers2016,JCP}.
The micro-to-macro transition for the cohesive traction vector
${}^0\vec{t}$ is given by
\begin{align}
    {}^0\vec{t}=\frac{1}{|\Theta_0|}\bigg[\int_{\Theta_0}\bs{F}\,{}^1\bs{S}\,\text{d}\Theta_0\bigg]\cdot {}^0\vec{N}.\label{eq:trac_pde}
\end{align}

As presented for interfaces in the work of Matou\v{s} et al.~\cite{matouvs2008multiscale}, we can use two sets of microscale
boundary conditions
\begin{align}     
    {}^1\vec{u}=\vec{0}\,\,\text{on  }\partial \Gamma_0^{\pm},\quad{}^1\vec{u}^+-{}^1\vec{u}^-&=\vec{0}\,\,\text{on }\partial\Theta_0^\pm,\quad {}^1\vec{t}^++{}^1\vec{t}^-=\vec{0}\,\,\text{on  }\partial \Theta_0^\pm,\label{eq:PDE}\\
    {}^1\vec{u}&=\vec{0}\quad\forall\,\vec{Y}\in\Theta_0,\label{eq:Taylor} 
\end{align}
which allows us to introduce a seamless model adaption at the
microscale. The typical semi-periodic boundary conditions, as given by
Eq.~(\ref{eq:PDE}), involve solving the microscale BVP of Eq.~(\ref{eq:pde_to_solve}) using a numerical method (e.g., finite
element method). This leads to high accuracy, but the computational
cost is unfortunately high. We will refer to this approach as the Full
Model (FM). Conversely, Eq.~(\ref{eq:Taylor}) describes the Taylor
Model (TM) which neglects fluctuations within $\Theta_0$, leading to a
relatively low computational cost in solving
Eq.~(\ref{eq:pde_to_solve}) due to its linear structure (i.e.,
summation). On the other hand, the accuracy of TM is lower than FM.
\subsection{Interface normal}
When dealing with adhesives that have complex geometries, it becomes
necessary to define the normal to the interface, ${}^0\vec{N}$ (see
Fig.~\ref{figure:potato_interface}). The macroscale
normal vector ${}^0\vec{N}$ needs to be perpendicular to the unit cell
top boundary. In this work, the rotation matrix is calculated using the
outer product, considering that the macroscale and microscale
coordinates share only one common basis vector, $\vec{X}_1=\vec{Y}_1$,
at a material point $\vec{x}\in\Gamma_0$. We can make this connection
between the macroscale and microscale basis vectors as we focus on statistically isotropic particulate systems. The rotation matrix
$\bs{R}$ is given by
\begin{equation}
  \bs{R} =
  \left(\begin{array}{c}
    \vec{Y}^*_1\\
    ^0\vec{N}\times{\vec{Y}}^*_1\\
    {}^0\vec{N}\\
    \end{array} \right),
\end{equation}
where $\vec{Y}^*=\bs{R}\vec{Y}$ is the new microscale point and
$\bs{R}^{\top}\bs{R}=\bs{1}$. In a discrete setting, we minimize the
difference between $\vec{X}_1$ and $\vec{Y}^*_1$ by projecting
\begin{equation}
\vec{Y}^*_1 = \vec{X}_1 -  \frac{\vec{X}_1\cdot{}^0\vec{N}}{||{}^0\vec{N}||}{}^0\vec{N},
\end{equation}
to eliminate possible drift due to the finite element discretization.
\subsection{Support vector regression for microscale model selection}
Selecting the appropriate microscale model on the fly, whether FM or
TM, is challenging when solving the coupled multiscale problem. The
interface, $\Gamma_0$, contains a large number of cohesive elements
and it is important to select FM or TM appropriately to retain the
required accuracy with minimal computational cost. Specifically,
it is desirable to adopt FM over areas of interest with large
deformation gradients and/or high stresses. On the other hand, TM can
be used in regions that are smooth.

Consequently, this process of selection effectively turns into a
statistical problem, characterized by assigning labels of $y=+1$ to FM
and $y=-1$ to TM. In statistical classification problems with two
classes, a decision boundary is built as a hypersurface that separates
the underlying vector space into two sets, one for each class (i.e.,
FM or TM). Decision boundaries can be clear-cut when the classes are
linearly separable, but they may be ambiguous in fuzzy logic-based
classification algorithms. An SVM finds the hyperplane with the maximum
margin~\cite{cortes1995support}, and the kernel trick is adopted to
turn the nonlinearly separable problem into a linearly separable
one. 

First, we define a SVM data point (i.e., a sample) as a jump
displacement $\vec{x}=\llbracket {}^0\vec{u}\rrbracket$ within the
cohesive element on $\Gamma_0$. Next, to retain the loading magnitude
$r=\|\llbracket {}^0\vec{u}\rrbracket\|$ over the Representative Unit
Cell (RUC), we introduce the jump vector components in a spherical
coordinate system 
\begin{align}
\llbracket {}^0u_1\rrbracket &= r\cos\phi\sin\theta,\label{eq:spherical_loads1}\\
\llbracket {}^0u_2\rrbracket &= r\sin\phi\sin\theta,\label{eq:spherical_loads2}\\
\llbracket {}^0u_3\rrbracket &= r\cos\theta,\label{eq:spherical_loads3} 
\end{align}
where $r$ is the radial distance, $\phi$ is the azimuthal angle, and
$\theta$ is the polar angle. Thus, when the magnitude of $r$ is
constant, the sampling space becomes 2D with
$\vec{x}=\left(\phi,\theta\right)$. 

To proceed, we consider $\varepsilon$-insensitive
SVR~\cite{drucker1996support} with the kernel trick for nonlinear
cases. In the binary SVM problem, we are given a training dataset
consisting of $N_t$ samples for each constant $r$ denoted as  
\begin{align}
\{(\vec{x}_i, y_i);\,i=1,2,\ldots,N_t\},\label{eq:n_training_sample}    
\end{align}
where each point is a two-dimensional real vector $\vec{x}_i\in
\mathbb{R}^2$ with labels $y_i=\pm1$ corresponding to FM and TM
simulations respectively. To create a nonlinear classifier, we adopt
one of the most commonly used Gaussian functions as a positive
semi-definite kernel 
\begin{align}
    k(\vec{x}_i,\vec{x}) = \exp{(-|\vec{x}_i-\vec{x}|^2)/\sigma^2},\label{eq:kernel}
\end{align}
where the depth of the kernel is denoted as $\sigma>0$, and $\vec{x}$
is a new data point to be classified. By introducing a dual set of
variables $\alpha_i$ and $\alpha_i^*$, the dual formula can be
expressed with an error tolerance of $\varepsilon$ that leads to 
\begin{align}
    \max_{\alpha_i,\alpha_i^*}\,\,\,\,&-\frac{1}{2}\sum_{i=1}^{N_t}\sum_{j=1}^{N_t}(\alpha_i-\alpha_i^*)(\alpha_j-\alpha_j^*)k(\vec{x}_i,\vec{x}_j)-\varepsilon \sum_{i=1}^{N_t}(\alpha_i+\alpha_i^*)+\sum_{i=1}^{N_t} y_i(\alpha_i-\alpha_i^*),\label{eq::svm_math_1}\\
    \text{subject to}&\quad\hspace{15mm} \sum_{i=1}^{N_t}(\alpha_i-\alpha_i^*)=0,\,\,\,
    0\leq \alpha_i\leq C,\,\,\,
    0\leq \alpha_i^*\leq C,
\end{align}
where $C$ is a positive constant. The sample $\vec{x}_i$ becomes
a Support Vector (SV) when $\alpha_i-\alpha_i^*\neq 0$. Note that the
Karush-Kuhn-Tucker complementarity conditions exist to explain
violation tolerance and the SV vectors can be
evaluated~\cite{deka2014support,fletcher2000practical}. This quadratic
programming problem is solved by a sequential minimal optimization
algorithm~\cite{fan2005working}. Data training can be seen as the
process of determining a regression function denoted by $f(\vec{x})$,
often referred to as the score function. Ultimately, we obtain the
nonlinear SVM score function for any newly added data points $\vec{x}$, 
\begin{align}
    f(\vec{x})=\sum_{\vec{x}_i\in SV}^{N_t} (\alpha_i-\alpha_i^*)k(\vec{x}_i,\vec{x})+b, \label{eq::svm_math_2}
\end{align}
where $b$ represents the bias.
\section{Adaptive multiscale strategy}\label{sec:3}
The fundamental concept of the adaptive multiscale strategy is to introduce an offline database through SVR training to accelerate
multiscale simulations. Over the course of this training, we generate numerous
samples to measure modeling errors. Subsequently, we label each sample
as either FM or TM based on a user-defined error tolerance to build
the decision boundary. 

In detail, we train the nonlinear SVM classifiers utilizing RUCs under
various loading conditions, $\vec{x}=(\phi,\theta)$. Following this,
we set a range for the magnitude of $r$ introducing the maximum jump
displacement $\lambda=\max(\|\llbracket{}^0\vec{u}\rrbracket\|)$,  
\begin{align}
    0\leq r \leq \lambda.\label{eq221}
\end{align}
We uniformly discretize the span into $N_s$ segments to limit the
number of total training steps (i.e., $N_s\times N_t$). Therefore, we can represent a set of
radial distances as
$\left\{r_1,r_2,\cdots,r_{N_s}\right\}=\sfrac{\lambda}{N_s}\left\{1,\,2,\,\cdots,\,N_s\right\}$. Each
radial distance covers an equal-sized segment (e.g., $0\leq r_1 \leq
\sfrac{\lambda}{N_s}$ and $\sfrac{\lambda}{N_s}< r_2 \leq
\sfrac{2\lambda}{N_s}$). Finally, we generate an $i$-th training
sample for each radial distance, $\vec{x}_i=(\phi_i,\theta_i)$, satisfying
\begin{align}
    \phi_i&= 2\pi H^i_{1},\label{eq:sample_1}\\
    \theta_i&= \pi H^i_{2},\label{eq:sample_2}
\end{align}
where $H^i$ is the $i$-th member of 2D Halton
sequence~\cite{halton1960efficiency}, with subscripts $(\bullet)_{1}$
and $(\bullet)_{2}$ for the first and second bases, respectively. In
this work, this quasi-random sequence skips the first 1000 values and
retains every 101st point followed by a reverse-radix scrambling
algorithm~\cite{kocis1997computational}. This approach ensures that
any dense clusters created by naive random sampling are removed and
facilitates easier implementation of low-discrepancy
sequences~\cite{wang2000randomized}.

For each sample generated, we obtain a solution from
Eq.~(\ref{eq:pde_to_solve}) for two sets of microscale boundary
conditions (i.e., FM and TM) in Eqs.~(\ref{eq:PDE})
and~(\ref{eq:Taylor}). Then, we establish a metric to evaluate the
modeling errors, each with subscripts $(\bullet)_{FM}$ and
$(\bullet)_{TM}$ to denote FM and TM, respectively. In this work, the
error metric is based on the traction vector
\begin{align}
    \mathcal{E}_{{}^0\vec{t}}&=\frac{\|{}^0\vec{t}_{FM}-{}^0\vec{t}_{TM}\|}{\|{}^0\vec{t}_{FM}\|}\label{eq:errordatabase}.
\end{align}

By introducing a user-defined modeling tolerance $\gamma$, we choose
the modeling error for each sample to ensure that $\mathcal{E}_{{}^0\vec{t}}
<\gamma$ when labeling the training samples $\vec{x}_i$ with TM. Otherwise, we use FM. In
the special case of unknown modeling error that violates
Eq.~(\ref{eq221}), all training samples are considered FM. As a
consequence of this labeling process, we assemble an offline database,
$\mathcal{DB}$, as a collection of trained score functions with the
nonlinear decision boundaries 
\begin{align}
\mathcal{DB} = \left\{f_1(\vec{x}), f_2(\vec{x}), \ldots, f_{N_s}(\vec{x})\right\}.\label{eq:offdata}
\end{align}
\section{Adaptive and parallel multiscale network}\label{sec:4}
\begin{figure}[!htb]
\centering
\includegraphics[width=0.90\textwidth]{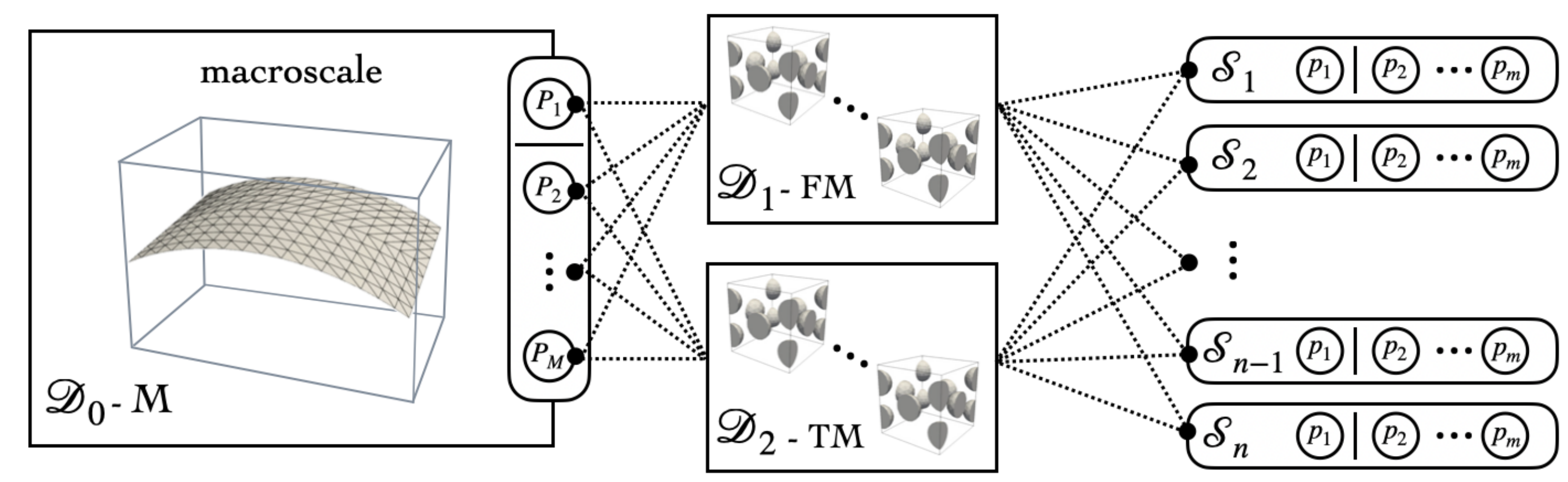}
\caption{Dependency trees of the adaptive multiscale network and parallel communication patterns.}
\label{figure:concept_adaptive}
\end{figure}
In a highly parallel computational environment managing two microscale
models emerges as the key challenge within the adaptive multiscale
modeling. This complexity arises because each microscale model (i.e.,
$\mathcal{D}_1-$FM and $\mathcal{D}_2-$TM) demands an additional
network abstraction layer in the dependency trees as shown in
Fig.~\ref{figure:concept_adaptive}. Moreover, the dependency trees
necessitate inter-object communications (shown by the dashed
lines). The macroscopic computational domain $\mathcal{D}_0$ assigns
one job for each cohesive element at the interface $\Gamma_0$ (see
Fig.~\ref{figure:potato_interface}). The microscale model adaption
(i.e., FM and TM), triggered by $\mathcal{D}_0$ jobs, leads to the
dynamic evolution of computational domains $\mathcal{D}_i$. These
domains are mapped to the microscale servers $\mathcal{S}_i$ requiring
load balancing. Subsequently, virtual services (i.e., $\mathcal{D}_i$
mapping to $\mathcal{S}_i$ and load balancing) call for proper
partitioning and management of computational resources. 

\begin{figure}[!htb]
\centering
\includegraphics[width=0.90\textwidth]{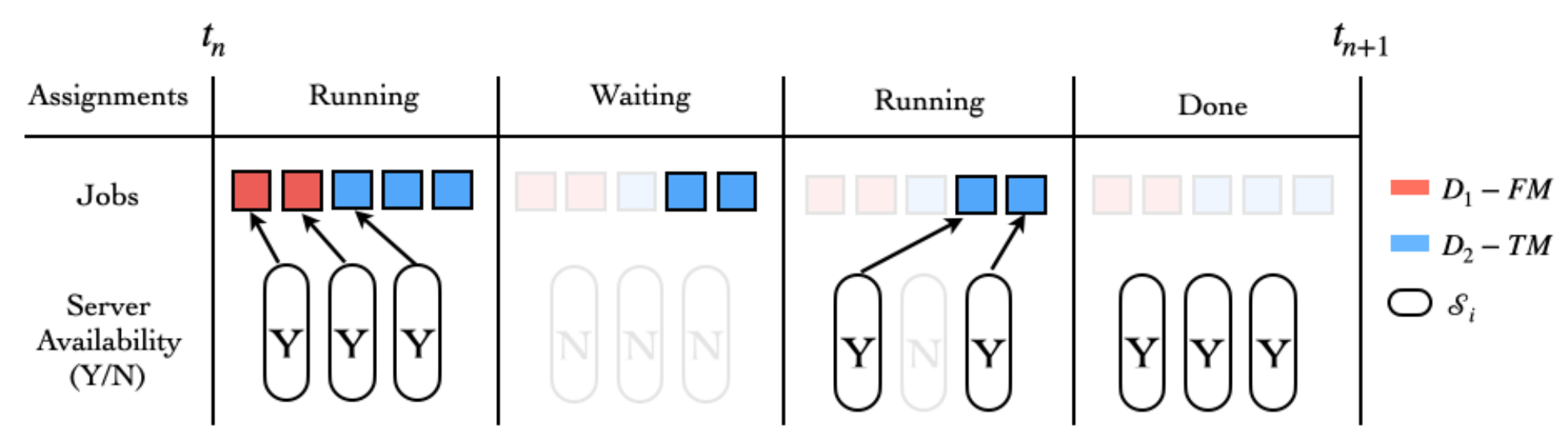}
\caption{Schematic of the workload balance through inter-domain communication for computational jobs. In this example, five jobs are processed by three servers between time steps $t_n$ and $t_{n+1}$. \label{figure::reallocation}}
\end{figure}
In this work, we present a novel multiscale network library --
\textit{multiscale\_net} -- that allows for uninterrupted dynamic
microscale model selection in a parallel
setting. \textit{multiscale\_net} is written in object-oriented C$++$
using the Message Passing Interface (MPI). New communicators (COMM)
are designed to manage the computational domains $\mathcal{D}_i$. To
distinguish between the three computations (i.e., macroscale, FM and TM
microscales), the generalized launch specification assigns each
numerical object (i.e., macro\-scale, FM, and TM) to a corresponding
domain (i.e., $\mathcal{D}_0-$M, $\mathcal{D}_1-$FM, and
$\mathcal{D}_2-$TM). The macroscopic domain $\mathcal{D}_0$ is
executed on a set of processors ($P_1,\cdots,P_M$) where $P_1$ denotes
the master processor, and $P_2,\cdots,P_M$ are worker processor. In
addition, $\mathcal{D}_0$ uses inter-domain communications to link the
microscopic domains $\mathcal{D}_i$ to servers $\mathcal{S}_i$
resulting in mappings $\mathcal{D}_1(\mathcal{S}_i)$ and
$\mathcal{D}_2(\mathcal{S}_i)$. Each of the macroscale processors
(i.e., $P_i$ in $\mathcal{D}_0$) communicates with the master
processor of each microscale server (i.e., $p_1$ in $\mathcal{S}_i$)
with $p_2,\cdots,p_m$ being worker processors. The typical
master-worker relation is established through intra-group
communications within each processor set (i.e.,
$\mathcal{S}_i$). Furthermore, due to the computational load imbalance
between FM and TM, domains $\mathcal{D}_i$ are dynamically reassigned
to microscale servers $\mathcal{S}_j$ based on the largest first
algorithm~\cite{leighton1979graph}. To achieve this, the inter-group
communication between microscale servers is restricted to processors
sharing the same task IDs, which leads to $p_i$ in $S_j$ communicating
with $p_i$ in $\mathcal{S}_k$, $\forall i=2,\cdots,m$. This simplifies
RUC data migration on available microscale servers, and reduces the
complexity of communication patterns. The procedure for balancing the
workload between time steps $t_n$ and $t_{n+1}$ is summarized in
Fig.~\ref{figure::reallocation}. 
\section{Numerical implementation and constitutive models}\label{sec:5}
The numerical implementation using FEM has been described in
\cite{MOSBY2014, MOSBY2015-1}. The weak forms of the macroscale and
microscale boundary value problems, Eq.~(\ref{eq:pde_to_solve_0}) and
Eq.~(\ref{eq:pde_to_solve}), are derived from a standard variational
procedure and we do not repeat them here for brevity. We implement SVM
based adaption as well as $multiscale\_net$ into our in-house
massively parallel library
$PGFem3D$\footnote{\url{https://github.com/C-SWARM/pgfem-3d}} for both
single-scale and multiscale simulations. The $PGFem3D$ solver has been
rigorously verified and validated in several previous
studies~\cite{lee2021numerical,MOSBY2014,MOSBY2015-1,MOSBY2015-2,Subber2018}. For
mesh generation, we employ $T3D$ meshing
tool~\cite{rypl1998sequential,em2002a} and create linear tetrahedral meshes
for both single-scale and multiscale simulations. The remaining task
involves specifying the constitutive models at the macroscale, Eq.~(\ref{eq:pde_to_solve_0}), and at the microscale, Eq.~(\ref{eq:pde_to_solve}).

At the macroscale, we use the typical Neo-Hookean solid 
\begin{align}
    {}^0W(\hat{{}^0\bs{C}})&=\frac{{}^0\mu}{2}[\text{tr}({}^0\hat{\bs{C}})-3],\label{eq:macro_strain_dens_dev}\\
    {}^0U({}^0J)&=\frac{{}^0\kappa}{2}[\text{exp}({}^0J-1)-\text{ln}({}^0J)-1],\label{eq:macro_strain_dens_vol}
\end{align}
where ${}^0\hat{W}({}^0\boldsymbol{C})$ represents the deviatoric and
${}^0U({}^0J)$ the volumetric component of the
macroscale strain energy density function. The macroscale Jacobian
is defined as ${}^0J = \text{det}({}^0\boldsymbol{F})$. Also,
${}^0\mu$ denotes the macroscale shear modulus, ${}^0\kappa$ is the
macroscale bulk modulus, and
${}^0\hat{\bs{C}}={}^0J^{-\sfrac{2}{3}}\,{}^0\bs{C}$ represents the
macroscale deviatoric right Cauchy-Green deformation tensor.

At the microscale, we use the continuum damage model developed by
Lee et al.~\cite{lee2021numerical}, which relies on the split
of the strain energy density function and the damage variable  
\begin{equation}
{}^1W(\boldsymbol{C}, \omega^d, \omega^v) = (1 - \omega^d) {}^1\hat{W}(\boldsymbol{C}) + (1 - \omega^v) {}^1U(J),
\end{equation}
where $\omega^d\in [0,1)$ and $\omega^v\in [0,1)$ are the deviatoric
and volumetric damage variables, respectively. The total damage
variable is defined by $\omega=\sqrt{(\omega^d)^2+(\omega^v)^2}/\sqrt{2}$. Similar to the
macroscale, we use the typical Neo-Hookean strain energy density
functions ${}^1W(\hat{\bs{C}})$ and ${}^1U(J)$ (see
Eqs.\eqref{eq:macro_strain_dens_dev} and
\eqref{eq:macro_strain_dens_vol}) along with the corresponding
deformation tensor $\bs{C}$, Jacobian $J$, and microscale material
parameters $\mu$ and $\kappa$.

The microscale PK2 stress is derived from the microscale strain energy densities as
\begin{align}
    {}^1\bs{S}=(1-\omega^d)2\frac{\partial {}^1\hat{W}}{\partial \bs{C}}+(1-\omega^v)\frac{\partial {}^1U}{\partial J}J\bs{C}^{-1}.
\end{align}
The split damage model is energy-driven and the evolution equations
for $\dot{\omega}^d$ and $\dot{\omega}^v$ are derived from the damage
energy release rates given by
\begin{align}
    Y^\bullet ({}^1\hat{W},{}^1U)=\alpha^\bullet\, {}^1\hat{W}+\beta^\bullet\, {}^1U,
\end{align}
where $()^\bullet$ denotes either the deviatoric $()^d$ or the
volumetric $()^v$ contribution. $\alpha^d,\alpha^v\in[0,1]$ and
$\beta^d,\beta^v\in[0,1]$ are material parameters controlling the
degree of volumetric and deviatoric coupling. We use
$\alpha^d=\alpha^v=1\;\forall J$, $\beta^d=\beta^v=1\;\forall J\ge 1$ and
$\beta^d=\beta^v=0\;\forall J<1$, to prevent pure volumetric damage
under compression. More details are provided in Lee et al.~\cite{lee2021numerical}.
\section{Numerical examples}\label{sec:6}
In this section, we present two numerical studies using our novel
adaptive multiscale strategy. The first verification study is designed
to assess the functionality of our adaptive approach. In this example,
we demonstrate progressive failure using a DCB with a curved
interface. Next, we simulate a large and industrially relevant
wind turbine blade application. In particular, we use the National
Rotor Testbed~\cite{gage2021laboratory} that reports on the
failure of a 13 meter-long wind turbine blade under ultimate static test.
\subsection{Double-cantilever beam}\label{DCB_section}
\begin{figure}[!htb]
\centering
\subfigure[]{\includegraphics[height=2.4in]{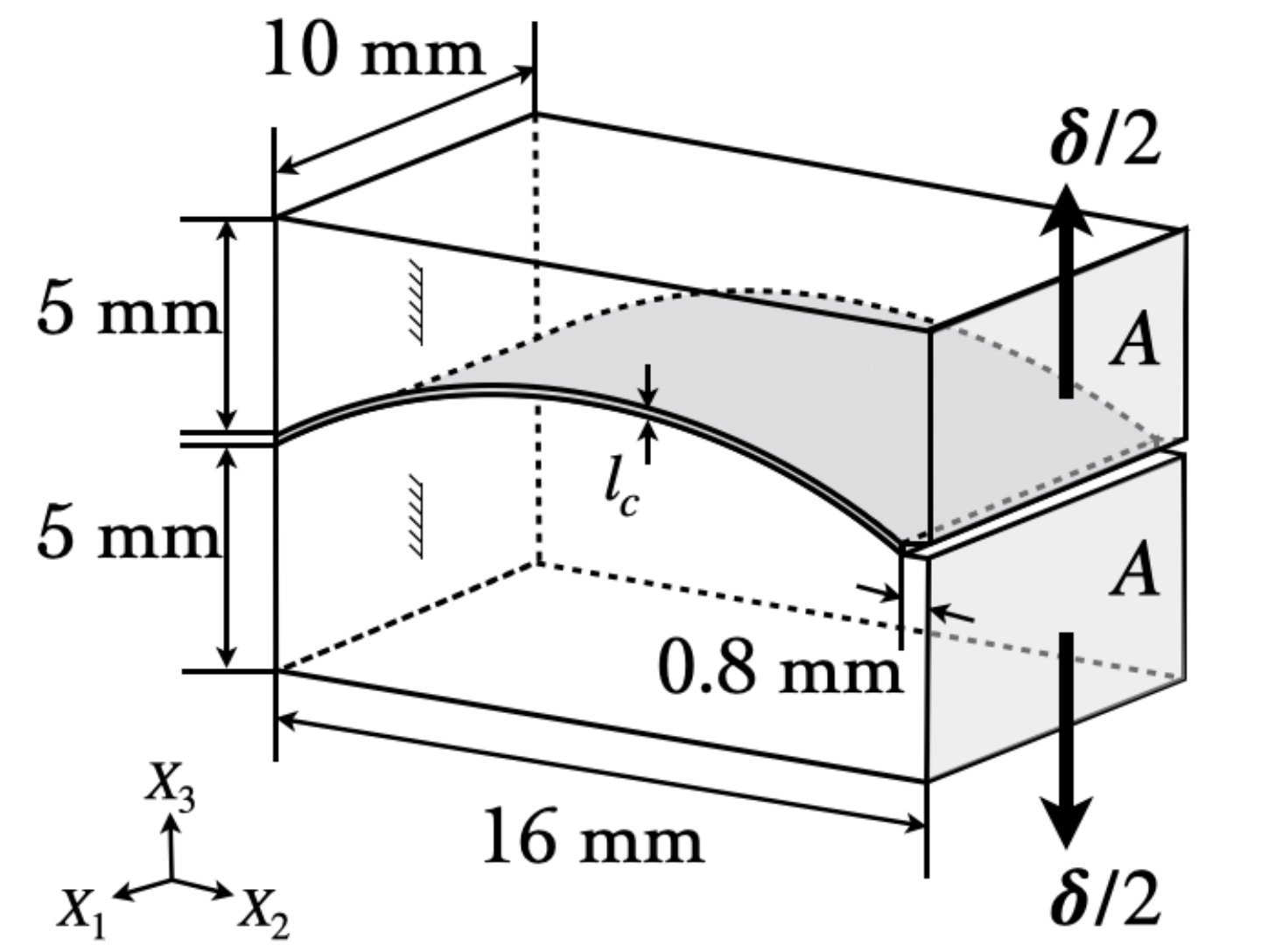}\label{fig:curve_macro_dcba}}\hspace{10mm}
\subfigure[]{\includegraphics[height=2.4in]{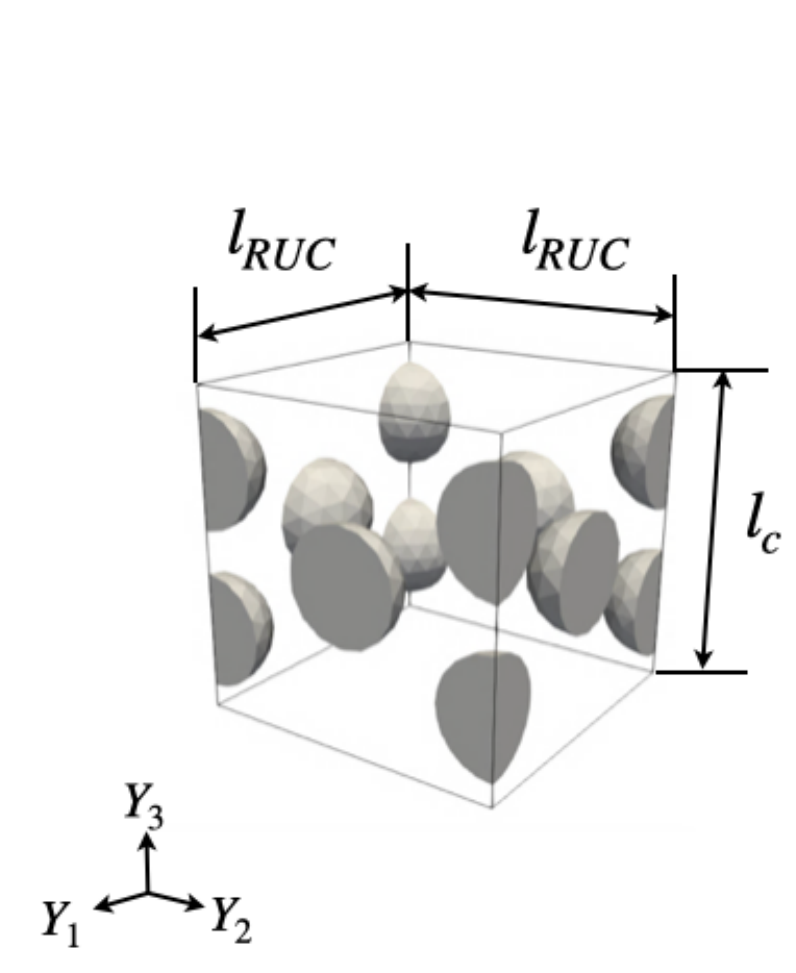}\label{fig:curve_macro_dcbb}}
\caption{Schematics of the multiscale DCB test. (a) Dimensions and boundary conditions, (b) RUC at the adhesive layers.}
\end{figure}

At the macroscale, two adherends are bonded by a curved adhesive with
an initial crack length of 0.8 mm, as shown in
Fig.~\ref{fig:curve_macro_dcba}. Cubic B\'{e}zier curves parametrize the
geometric arc of the adhesive with a central angle of $\pi/3$
radians. Clamped boundary conditions are applied to the left end of the DCB. Then, we impose a displacement load of $\delta = 0.02$ mm,
distributed at a rate of 3 mm/min to the surface $A$ of each
block, ensuring quasi-static loading conditions. The surface $A$ is
constrained in the $X_2$ directions and remains planar during the loading
history. Additionally, the reaction force $F$ acting on the surface
$A$ is computed for analysis.

The macroscale material properties are characterized by Young's
modulus ${}^0E = 210 \times 10^3 \text{ MPa}$, shear modulus ${}^0\mu
=72 \times 10^3 \text{ MPa}$, bulk modulus ${}^0\kappa=167 \times 10^3
\text{ MPa}$, and Poisson's ratio ${}^0\nu = 0.29$. The macroscale
mesh consists of 3,210 nodes, 14,609 linear tetrahedral finite
elements, 320 cohesive elements, and 8,860 Degrees of Freedom
(DOFs). This configuration is determined through the mesh refinement
study using FM with a varying number of cohesive elements (i.e., from
192 to 572).

In the next step, we attach 320 RUCs to the material points on the
adhesive surface, with each RUC paired with a corresponding cohesive
element. For the RUCs, we adopt the material properties of the
surrogate model detailed in~\cite{MOSBY2015-1,lee2021numerical}, which
consists of stiff nylon particles~\cite{luna2022influence} embedded in
a polyurethane structural adhesive~\cite{di2020structural} as shown in
Fig.~\ref{fig:curve_macro_dcbb}. Each RUC measures
$100\times100\times100\,\mu\text{m}^3$, and includes 4 particles with
a diameter of 18.2 $\mu$m resulting in a particle volume fraction of
$c_p=10.10\,\%$. Particle packing is arranged to maintain axial
symmetry about the $Y_3$ axis in this verification example. The
thickness of the adhesive layer is denoted as
$l_c=100\,\mu\text{m}$. The size of each RUC, $l_{RUC}=100\;\mu$m, is
determined through independent numerical simulations as
in~\cite{MOSBY2015-1}. The macroscopic loading rate is restricted to
$\|\llbracket{}^0\dot{\vec{u}}\,\rrbracket/l_c\|\leq1.0\,\text{s}^{-1}$
due to the relatively low rate sensitivity of the adhesive
materials. The material properties of the split damage model (see Lee et al.~\cite{lee2021numerical}) are given in
Table~\ref{tab:material_prop}. We note that damage propagation does
not occur within the stiff nylon particles.
\begin{table}[!ht]
\fontsize{9}{11}\selectfont
\begin{center}
\caption{Material properties of the particles and matrix. $Y_{in}$ is the initial damage threshold, $p_1$ and $p_2$ are damage parameters, and $\mu^*$ denotes the damage viscosity.}\label{tab:material_prop}\vspace{2mm}
\begin{tabular}{p{1.3cm} |p{1.5cm} |p{1.5cm} |p{1.5cm} | p{0.8cm} | p{1.5cm} | p{0.8cm} | p{0.8cm} | p{1.2cm} }\hline
&$E$ [MPa] & $\mu^*$ [MPa] & $\kappa$ [MPa] & $\nu$ [-] & $Y_{in}$
  [MPa] & $p_1$ [-]& $p_2$ [-] & $\mu^*$ [s$^{-1}$] \\ \hline
Particle &$2.40\times 10^3$ & $8.96\times 10^2$ & $2.50\times10^3$ & 0.34 &- & - & - & -\\ \hline
Matrix &$8.00\times 10^2$ & $2.99\times 10^2$ & $8.33\times10^2$ & 0.34 &0.15 & 8.0 & 2.5 & 100.0\\ \hline
\end{tabular}
\end{center}
\end{table}

The RUC is discretized with 4,004 nodes, 23,344 finite elements,
and 11,586 DOFs with an average mesh size of $3.5\,\mu\text{m}$. This
configuration is obtained from a mesh convergence study solving
Eq.~(\ref{eq:pde_to_solve}), using uniaxial tensile and pure shear
loading conditions for FM. The average mesh size is progressively
adjusted to $3.5$, $2.8$, and $1.9$ $\mu$m, demonstrating solution
convergence. We further conduct tensile loading using FM to determine
the optimal size of each microscale server $\mathcal{S}_i$ (not
presented for brevity). This strong scaling study shows near-ideal
speedup with 4 cores. Thus, we allocate 4 cores to each $\mathcal{S}_i$. In
general, \textit{multiscale\_net} allows for non-uniform recourse
allocation. However, in this work, we assign the same number of cores
to each server as both FM and TM are locally parallelized.

Leveraging the chosen mesh size and resource allocation, we assign the
maximum jump displacement to be $\lambda=0.1\times l_c$ (see
Eq.~(\ref{eq221})) as the traction-separation relation predicts
complete material failure at 10\% elongation~\cite{MOSBY2015-1}. We
discretize this jump into $N_s=100$ segments and generate
$N_t\in[100,\,1000]$ training samples for each segment. Also, we exploit
the geometric symmetry of the RUC by substituting $\phi_i=
\frac{\pi}{4} H^i_{1}$ in Eq.~(\ref{eq:sample_1}) to assess the
density of $\mathcal{DB}$ (see Section~\ref{sec:3}). In the
computational environment, we employ job arrays utilizing a total of
272 cores to build $\mathcal{DB}$ in parallel.

\begin{figure}[!htb]
\centering
{\includegraphics[width=0.49\textwidth]{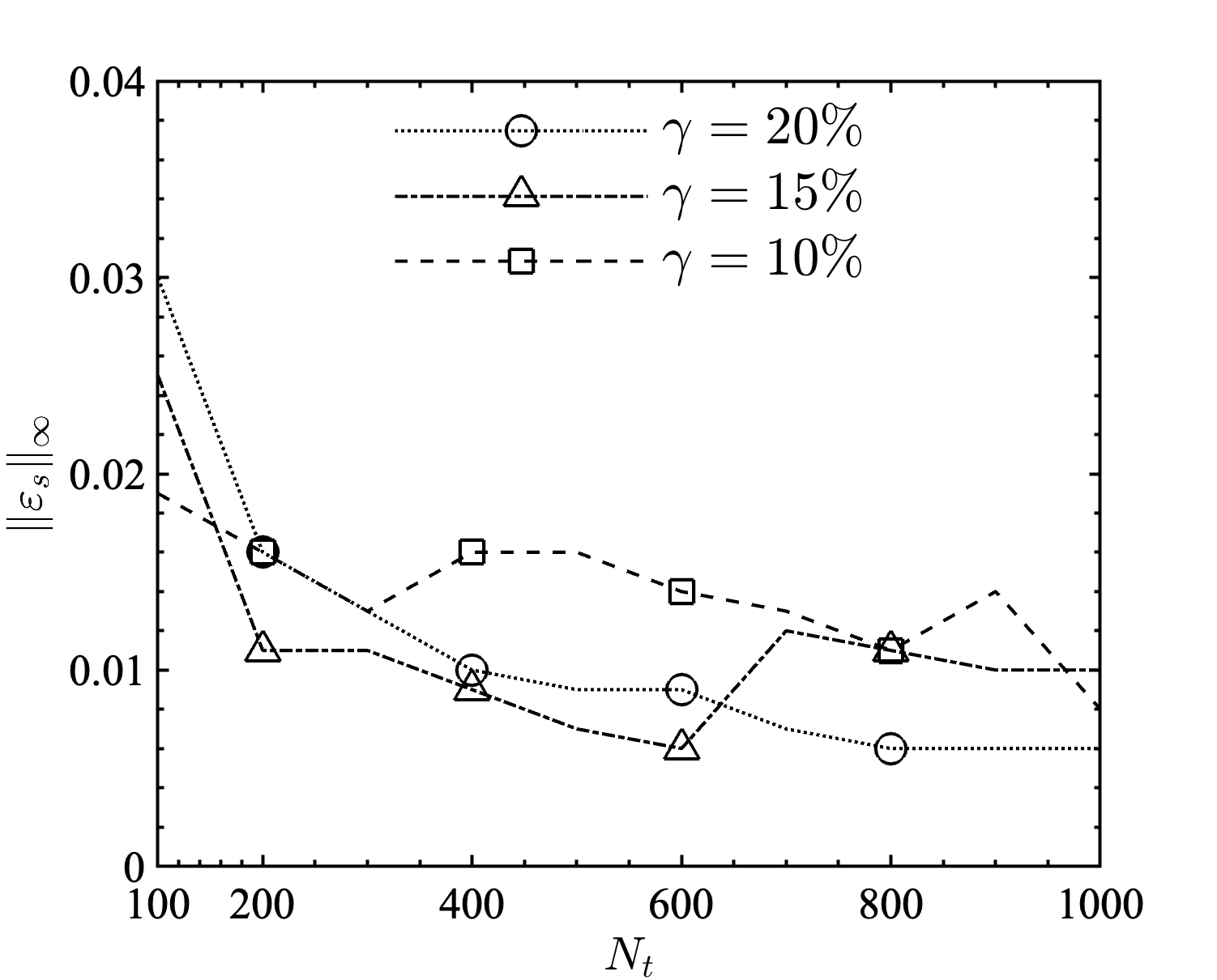}}
\caption{Classification error for nonlinear classifiers trained on data with varying model tolerances of $\gamma=10\%$, 15\%, and 20\%.\label{fig:classification_err_nylon}} 
\end{figure}
To address the impact of $N_{t}$ in Eq.~(\ref{eq:n_training_sample})
for each segment, $s$, on the accuracy of the SVM classifier, we
define the segment classification error
\begin{align}
\varepsilon_s = \frac{N^i}{N_{\text{C}}},\label{eq:classfication_err}
\end{align}
where $N^i$ denotes the number of incorrect classifications and
$N_C(\phi,\theta)$ represents the number of classifications (i.e.,
testing samples), which remains constant for each segment. For this
error analysis, we generated $N_C=1000$ quasi-random testing samples,
$(\phi,\theta)$, that are independent of the training samples, $N_t$.

Fig.~\ref{fig:classification_err_nylon} demonstrates the maximum
classification error for all segments $s$ during the loading history
up to the maximum displacement jump of $\lambda=0.1\times l_c$. We can
see that error drops quickly and remains relatively constant for a
database with $N_t\ge 500$ for the nonlinear SVM classifier with modeling
error tolerance $\gamma =10\%,\, 15\%,\, \text{and } 20\%$. For
all cases, we misclassified less than 1.8\% of tests. Based on these
observations, we set $N_t=500$ for the adaptive multiscale simulations
to follow. Consequently, we obtain a database
$\mathcal{DB}=\left\{f_1,f_2,\cdots,f_{N_s}\right\}$ by assessing
$N_s=100$ radial distances, $r$, based on Eq.~(\ref{eq221}). 

\begin{figure}[!htb]
\centering
\subfigure[]{\includegraphics[width=0.35\textwidth]{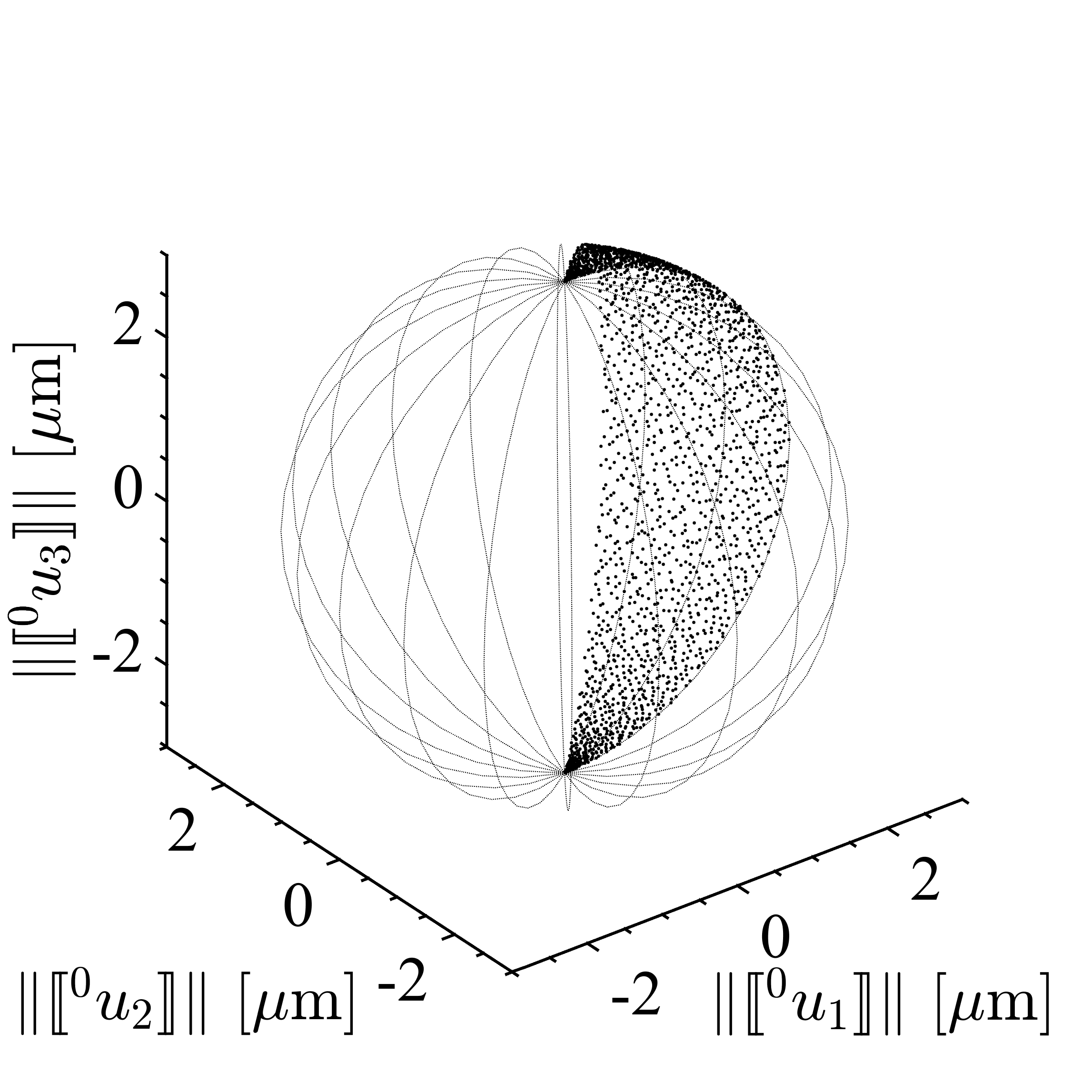}\label{svm_nylon_samplea}}
\subfigure[]{\includegraphics[width=0.60\textwidth]{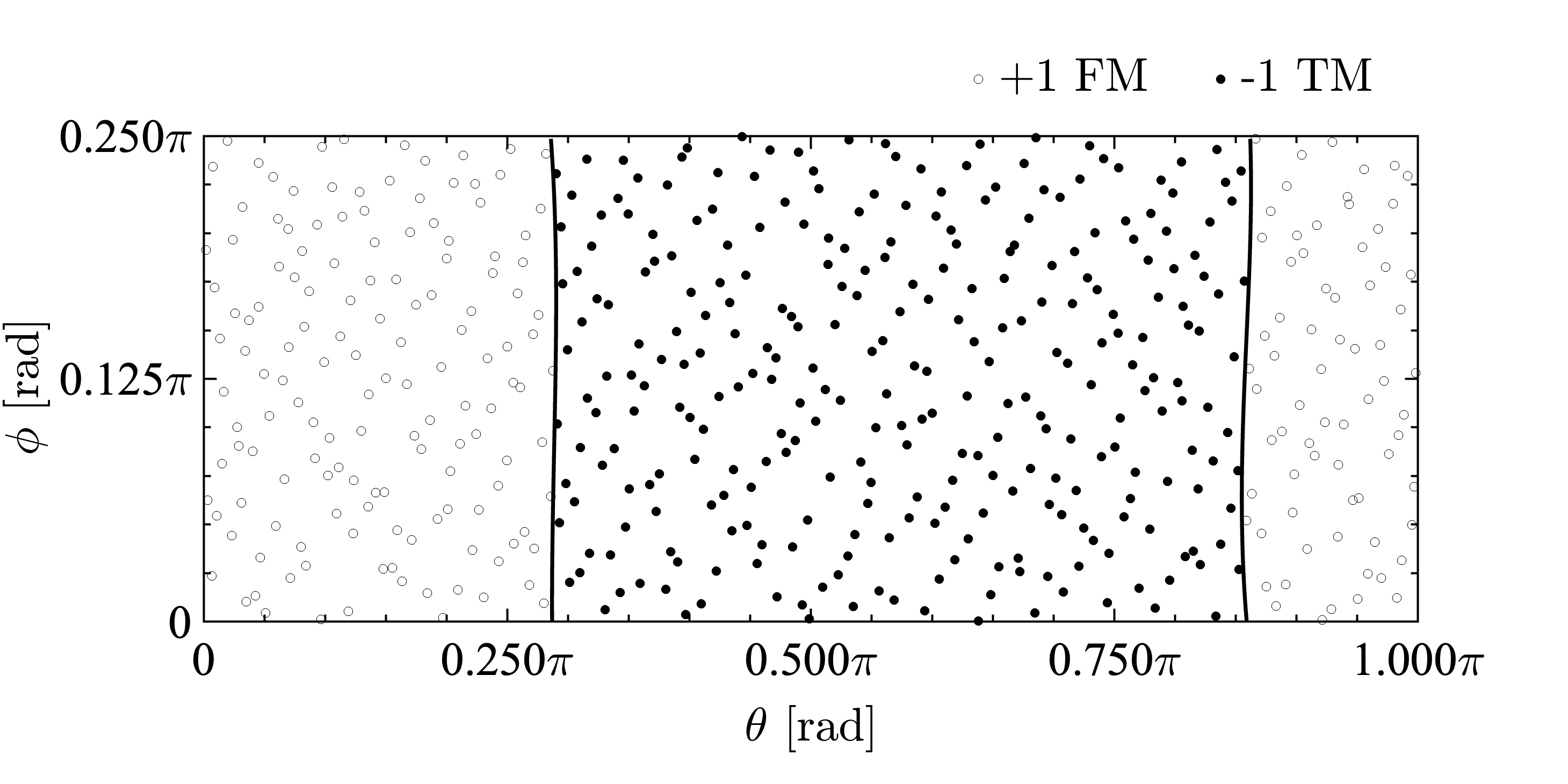}\label{svm_nylon_sampleb}}
\caption{An offline database trained on $N_t=500$ samples and $r=30$\% elongation with $\gamma=15$\% model tolerance. (a) 3D jump displacement space and (b) 2D spherical coordinate system.\label{svm_nylon_sample}} 
\end{figure}
Fig.~\ref{svm_nylon_sample} illustrates an offline database trained on
$N_t=500$ samples and $r=30$\% elongation (i.e., $r=0.3\lambda$) with
$\gamma=15$\% model tolerance. Fig.~\ref{svm_nylon_sampleb} shows the
decision boundary created by SVR in a 2D spherical coordinate system,
with dense clusters removed to ensure even sampling. Moreover, we note
that the less expensive TM covers a large range of loading
conditions. This illustrates the possible modeling acceleration by
adaptively selecting TM over FM with a user prescribed tolerance $\gamma$.

As an initial condition, adaptive multiscale simulation starts always using TM, adaptively changing to FM based on $\mathcal{DB}$. The
remaining figures present DCB results from adaptive multiscale
simulations using a total of 272 cores. We establish 64 microscale
servers $\mathcal{S}_i$, each with 4 cores per server, based on the
strong scaling study, with a total of 256 cores. This enables
\textit{multiscale\_net} to balance the work of the adaptive simulations as shown in Fig.~\ref{figure::reallocation}. The remaining 16 cores were
allocated to the macroscale domain $\mathcal{D}_0$.

\begin{figure}[!htb]
\centering
\subfigure[]
{\includegraphics[width=0.49\textwidth]{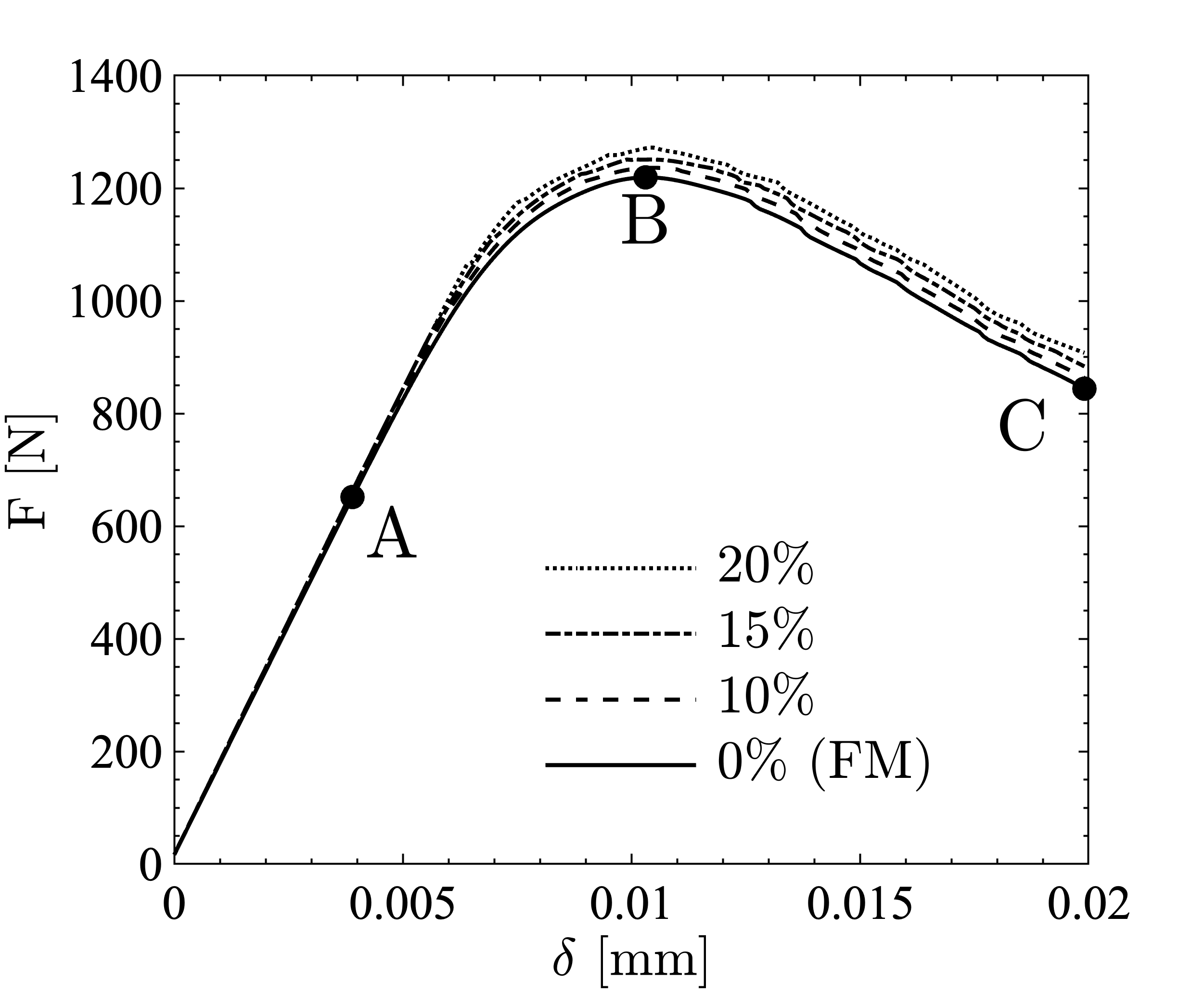}\label{fig:verification_result_alla}}
\subfigure[]
{\includegraphics[width=0.49\textwidth]{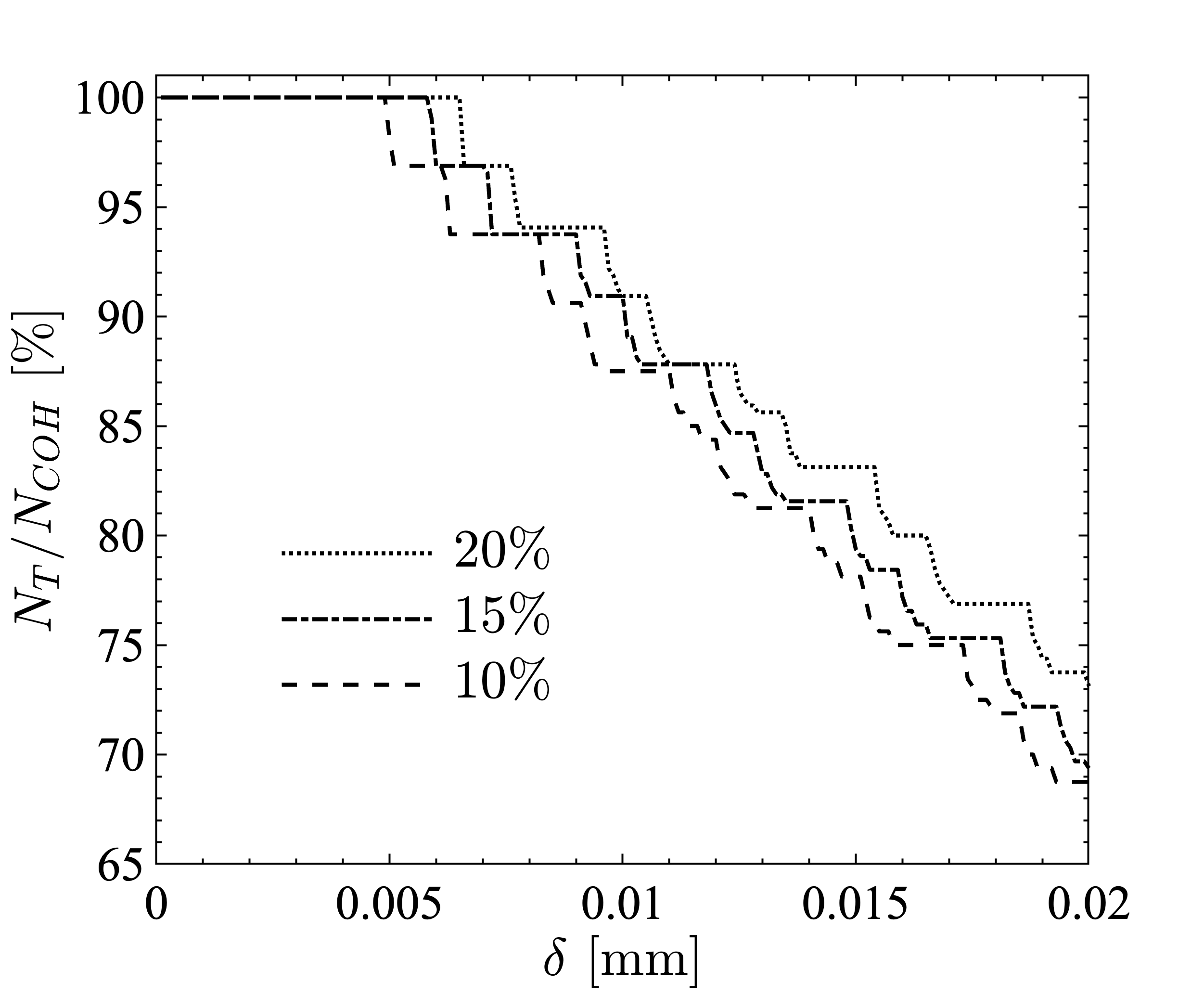}\label{fig:verification_result_allb}}
\caption{Adaptive multiscale results for the DCB test. (a) The $F$--$\delta$ response, and (b) the ratio of TM versus FM cohesive elements during the simulation.\label{fig:verification_result_all}} 
\end{figure}
\begin{figure}[!htb]
\centering
{\hspace{10mm}\includegraphics[width=0.85\textwidth]{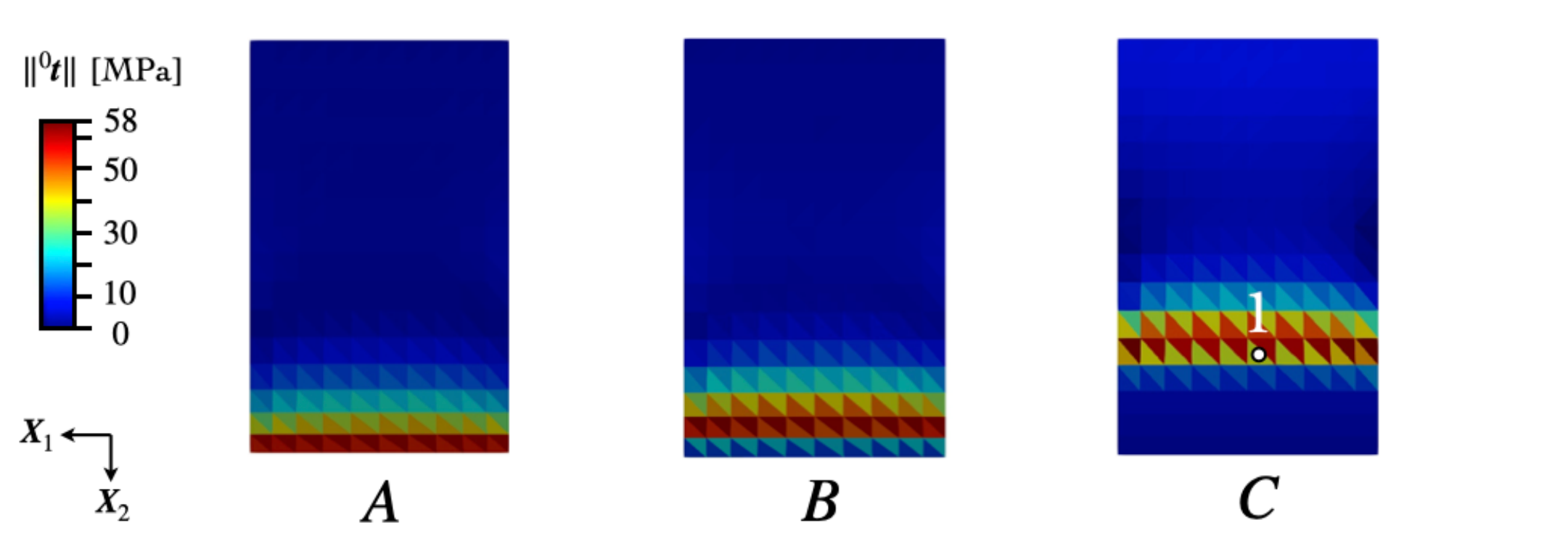}}
\caption{The magnitude of the cohesive traction vector for $\gamma=0\%$ (FM) at the points marked in Fig.~\ref{fig:verification_result_alla}.}\label{fig:verification_result_visual} 
\end{figure}

Fig.~\ref{fig:verification_result_alla} shows the $F-\delta$ response for
different $\gamma$ values. The difference in the peak load between FM
and each adaptive simulation is 1.40\% for $\gamma=10\%$, 2.65\% for
$\gamma=15\%$, and 4.38\% for $\gamma=20\%$, respectively. We note
that $\gamma=0\%$ represents the FM simulation. Over the loading
history, the adaptive strategy progressively adds more FM cells to
improve accuracy in the adhesive layer as damage in the RUCs
evolves. Fig.~\ref{fig:verification_result_allb} displays the percent
ratio of TM versus FM cohesive elements (i.e., RUCs) with different
$\gamma$ values. For the initial loading history, $\delta\lesssim
0.005$ mm, the cell response is hyper-elastic and the algorithm does
not adapt from TM. The adaptive simulation with the tighter $\gamma$
tolerance adapts earlier and progressively switches from TM to FM as
damage propagates in the adhesive layer. The stepwise nature of the adaption is due to the front propagating over cohesive elements at
the interface (see Fig.~\ref{fig:verification_result_visual}).

\begin{figure}[!htb]
\centering
\subfigure[]
{\includegraphics[width=0.25\textwidth]{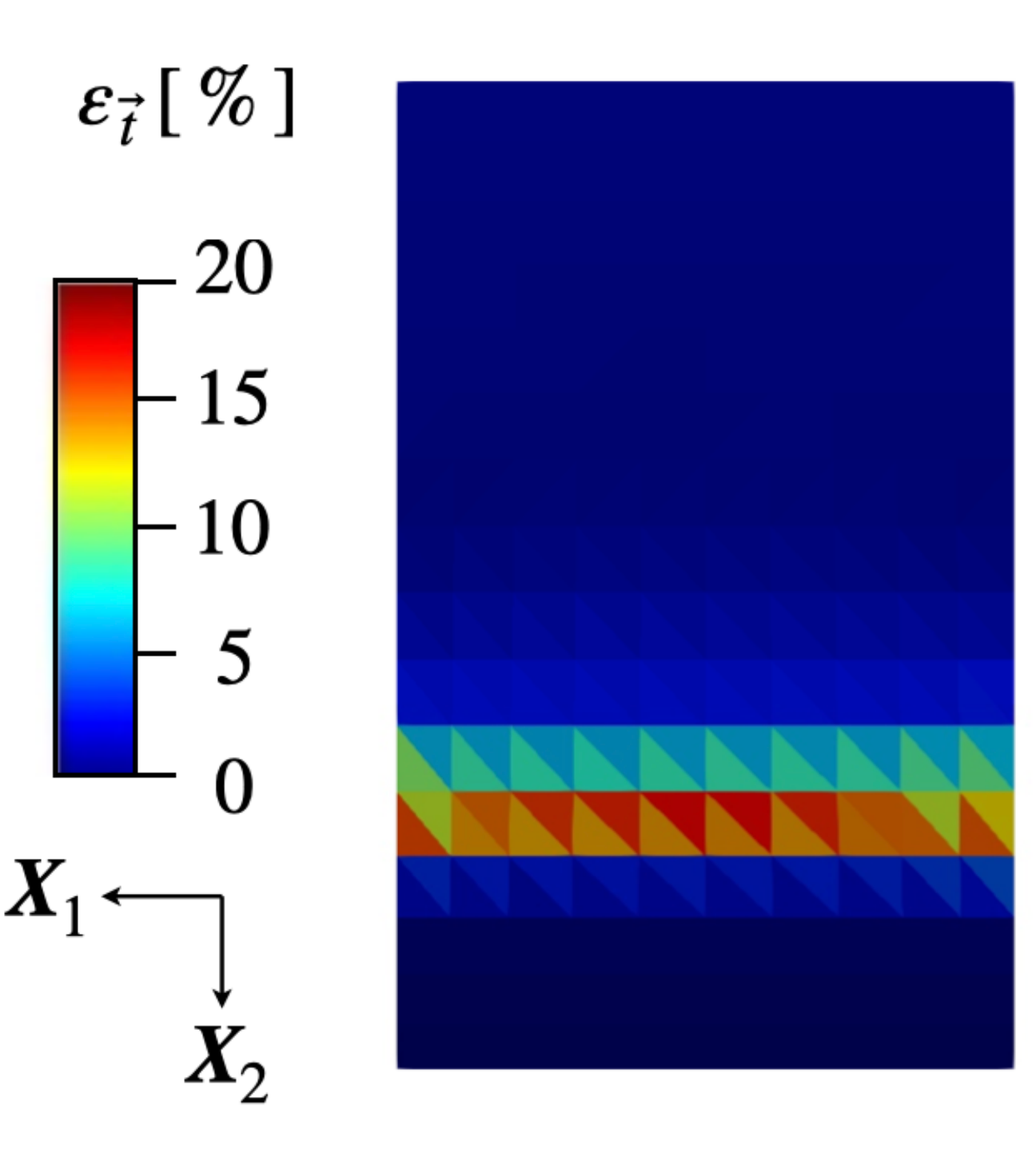}\label{fig:example1_comparisona}}\hspace{5mm}
\subfigure[]
{\includegraphics[width=0.25\textwidth]{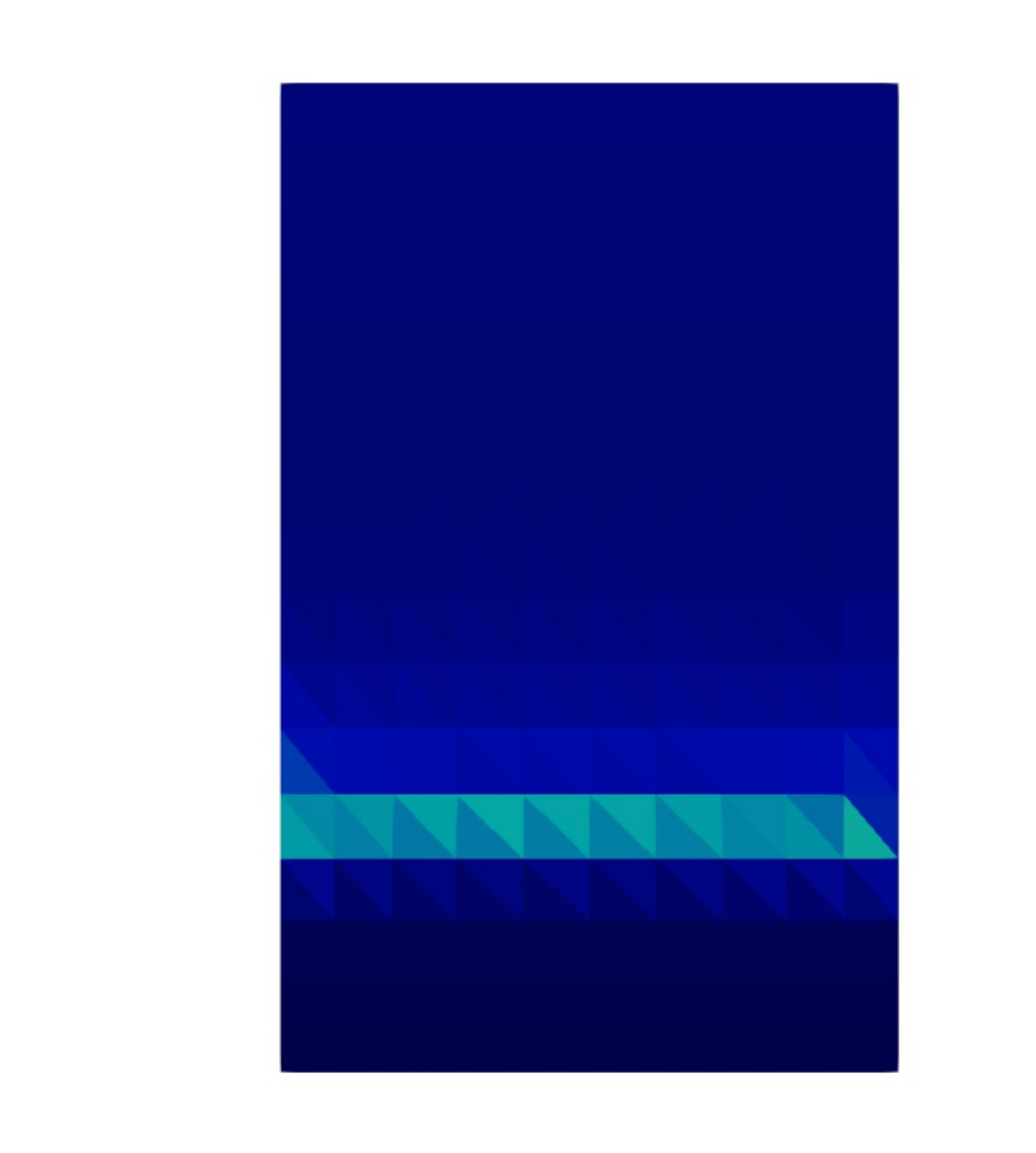}\label{fig:example1_comparisonb}}
\caption{The traction vector error at the end of the loading history for (a) $\gamma=15$\% and (b) $\gamma=10$\%.\label{fig:example1_comparison}}
\end{figure}

Fig.~\ref{fig:verification_result_visual} presents the magnitude of
the traction vectors, computed using Eq.~(\ref{eq:trac_pde}), at specific
points along the loading history identified in
Fig.~\ref{fig:verification_result_alla}. These points correspond to
the proportional limit (A), the peak load (B), and the progressive
damage state at the end of the simulation (C). To assess the impact of
modeling tolerance $\gamma$ on the adhesive response, we analyze the
macroscopic traction vector error
\begin{align}
    \varepsilon_{\vec{t}}&=\frac{\|{}^0\vec{t}_{FM}-{}^0\vec{t}\|}{\|{}^0\vec{t}_{FM}\|_\infty}\times 100\%.
\end{align}
Fig.~\ref{fig:example1_comparison} shows the traction vector error for
$\gamma=15\%$ and $\gamma=10\%$ values. The error drops precipitously
as the modeling tolerance approaches FM (i.e., full PDE solution). 

\begin{table}[!ht]
\fontsize{9}{14}\selectfont
\begin{center}
\caption{The speedup of adaptive multiscale simulations.}\label{tab:verification_result_all}\vspace{2mm}
\begin{tabular}{ p{1.0cm} | p{1.6cm} | p{1.6cm} } \hline
$\gamma$ [\%]  &$N_T/N_{COH}$  &  $T_{FM}/T_{AM}$ \\  \hline
20 & 0.731 & 1.55 \\ \hline
15 & 0.694 & 1.42 \\ \hline
10 & 0.488 & 1.33 \\ \hline
\end{tabular}
\end{center}
\end{table}
To assess the savings in simulation time, we compute the speedup
ratio $T_{FM}/T_{AM}$, where $T_{FM}$ represents the runtime of the
FM simulation, and $T_{AM}$ denotes the runtime of the adaptive
multiscale simulations, respectively. As expected, the large error
tolerance leads to large time savings, as depicted in
Table~\ref{tab:verification_result_all}. 

\begin{figure}[!htb]
\centering
\subfigure[]{\includegraphics[width=0.45\textwidth]{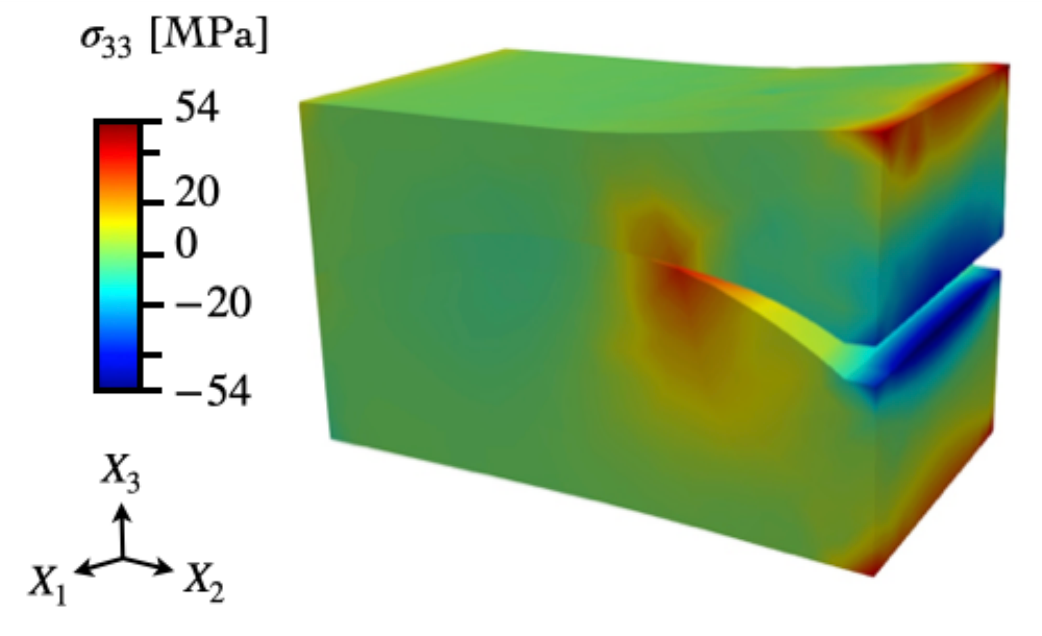}\label{figure:ams_result_2_dcb_macroa}}\hspace{5mm}
\subfigure[]{\includegraphics[width=0.45\textwidth]{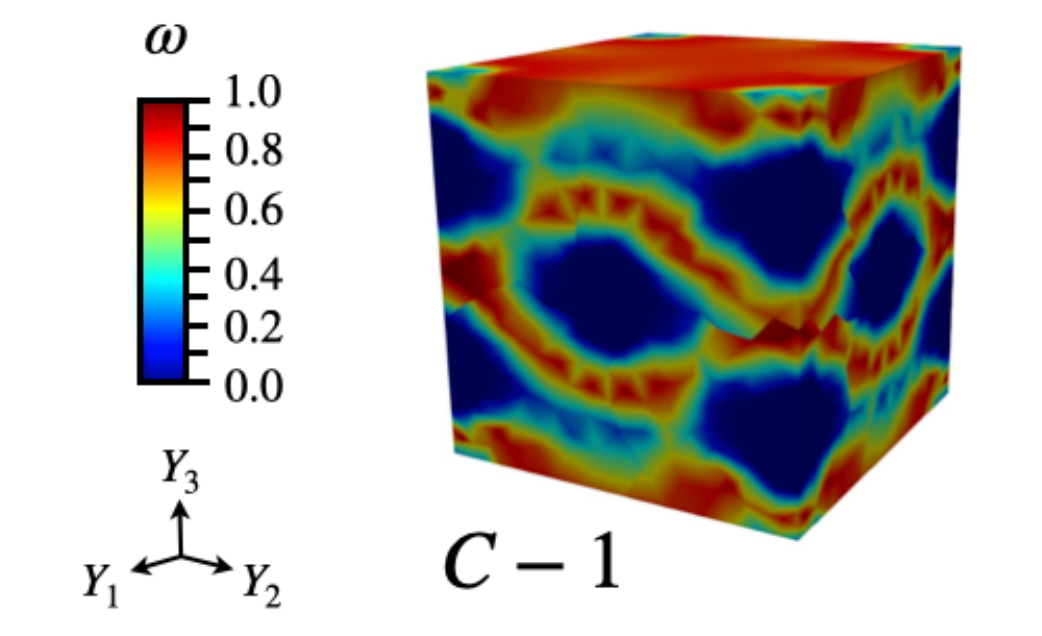}\label{figure:Cell_omega}}
\caption{The DCB macroscale and microscale results at point C in Fig.~\ref{fig:verification_result_alla}. (a) The stress component $\sigma_{33}$ in the deformed configuration (deformations magnified 50$\times$) and (b) damage pattern in the RUC. The cell location is depicted in Fig.~\ref{fig:verification_result_visual}.}\label{figure:ams_result_2_dcb_macro}
\end{figure}
\begin{figure}[!htb]
\centering
{\includegraphics[width=0.49\textwidth]{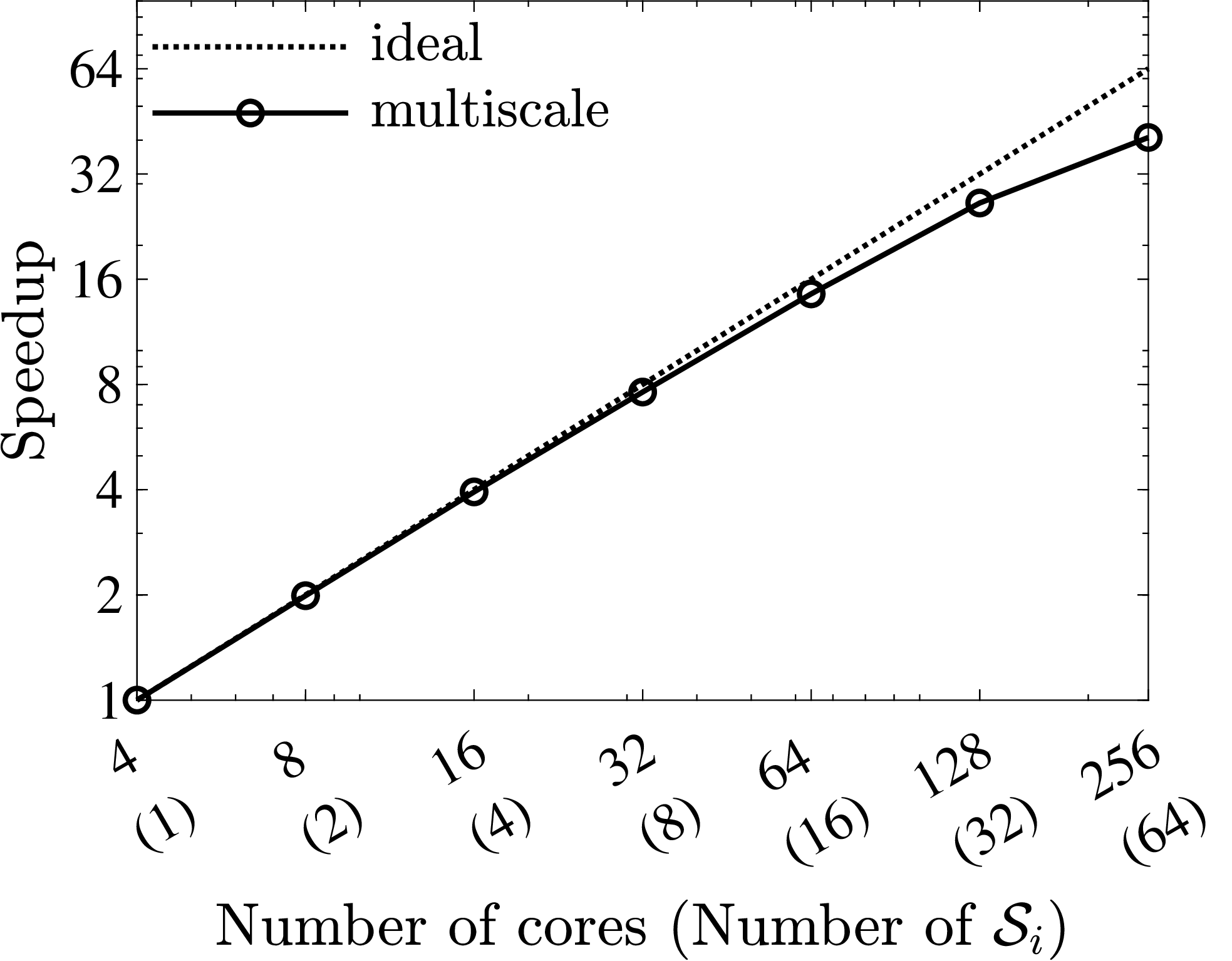}}
\caption{Strong scaling of the multiscale simulations.}\label{fig:speedup_dcb_parallel}
\end{figure}
Fig.~\ref{figure:ams_result_2_dcb_macroa} shows the 33 Cauchy stress
tensor component in the deformed configuration. There is a large
stress concentration at the crack tip. We note that stress
concentrations on the face $A$ of the beam are the result of the
prescribed boundary conditions. The face $A$ is not allowed to rotate
during the beam loading. Fig.~\ref{figure:Cell_omega} displays the
damage pattern in the microstructure at the end of the loading history
for the cell behind the crack front (see white marker $3$ in
Fig.~\ref{fig:verification_result_visual}). The cell failed
progressively around the particles and close to the top and bottom
adherends.

Fig.~\ref{fig:speedup_dcb_parallel} shows the strong scaling
achieved by the \textit{multiscale\_net} library for the multiscale
simulations with $\gamma=0$\%. The computational speedup is defined as
$\sfrac{t_n}{t_N}$, where $t_N$ is the time to compute the simulation
using $N$ resources (e.g., microscale servers, $\mathcal{S}_i$) and
$t_n$ is the time to compute the same simulation with more resources,
$n$. In our case, we start with one micro-server $\mathcal{S}_1$
consisting of 4 cores and increase the computational resources to 64
micro-servers with 256 cores in total. The \textit{multiscale\_net}
retains a near ideal $64\times$ speedup for the full computational
allocation.
\subsection{Ultimate static test of National Rotor Testbed}
In the second example, we model the geometry of a 13-meter NRT blade
that was tested by
NREL~\cite{gage2021laboratory,murray2019fusion,kelley2015aerodynamic}. In
particular, we aim to predict the ultimate bending moment that was
experimentally measured as 228.7 kN$\cdot$m. In this work, we assume
that ultimate failure starts at the adhesive layer. Thus, we
associate ultimate failure with fully failed RUCs, which indicates
crack propagation. Therefore, we compute the ultimate moment, $M^h$,
as a moment where the jump opening displacement
$\|\llbracket{}^0\vec{u}\,\rrbracket\|$ starts rapidly growing (i.e.,
the adhesive layer starts opening and the crack starts propagating).

\begin{figure}[!htb]
\centering
\subfigure[]{\includegraphics[width=0.72\textwidth]{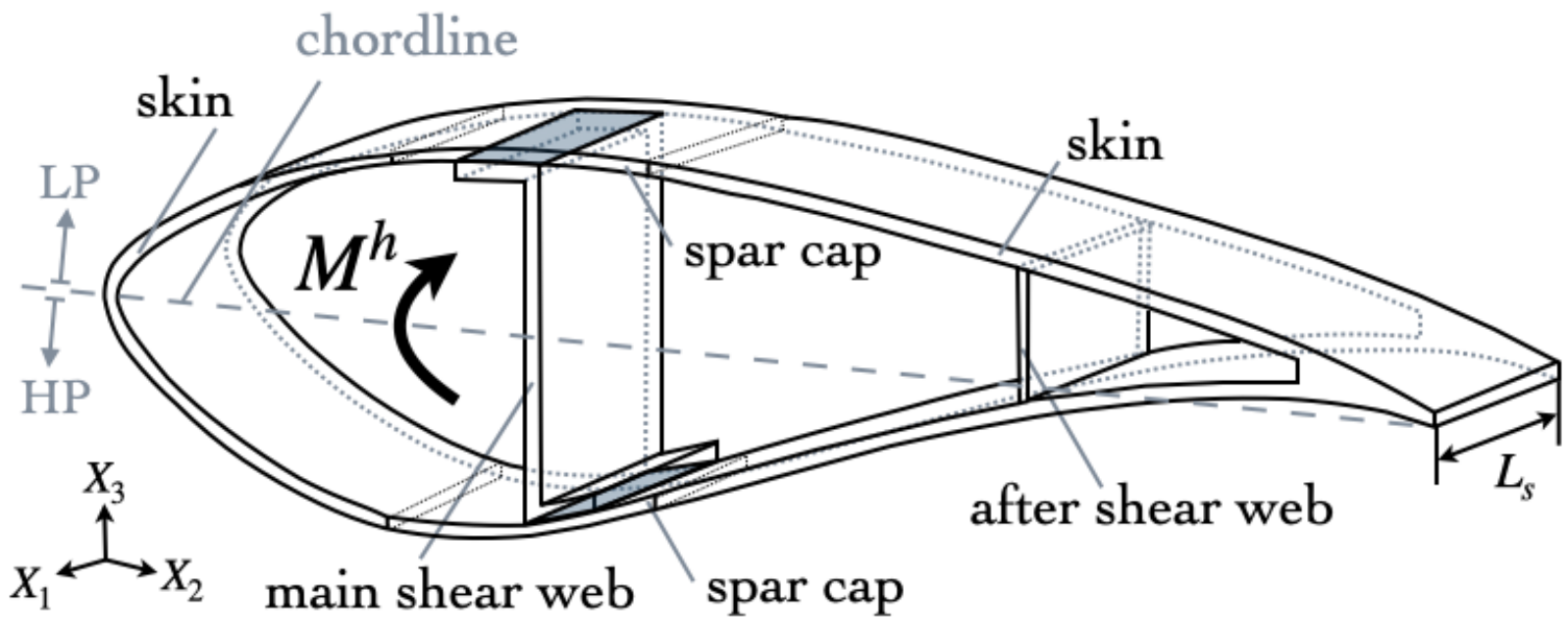}\label{fig::NRT_macroa}}
\subfigure[]{\includegraphics[width=0.23\textwidth]{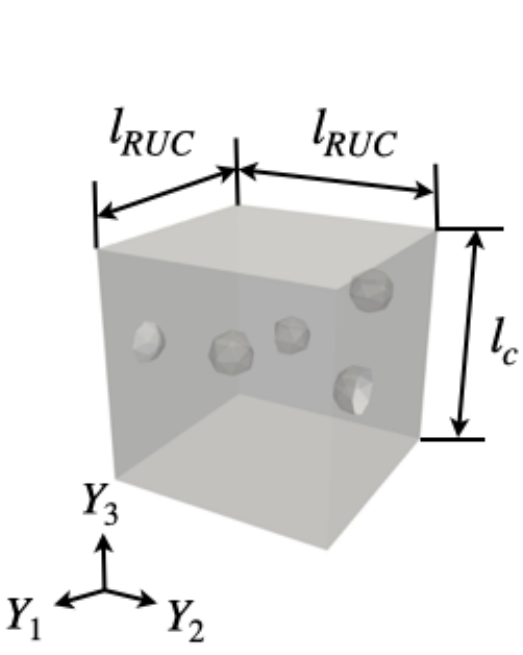}\label{fig::NRT_macrob}}
\caption{NRT wind turbine blade. (a) Geometry and boundary conditions. (b) RUC of the adhesive layer with manufacturing defects (i.e., voids)}\label{fig::NRT_macro}
\end{figure}
Fig.~\ref{fig::NRT_macroa} depicts the blade geometry, replicating the
S814 airfoil section with a chordline width of 1.450 m and a span-wise
length of $L_s=0.20$ m. We focus on the spanwise section from 2.7 to
2.9 meters, because the blade failed near the 2.9 m station as
reported in~\cite{gage2021laboratory}. The span-wise length of
$L_s=0.2$ m was selected based on the blade uniformity. Moreover, we
have performed independent numerical simulations to confirm that this
span-wise length does not influence the multiscale simulation results
(i.e., $L_s$ of 0.25 m and 0.20 m yield identical results). The
low-pressure side is denoted as LP and the high-pressure side as
HP.

The root of the blade (i.e., $X_1=0$) is clamped to simulate fixed
boundary conditions. We impose a bending moment $M^h$ about $X_2$ to
introduce a constant bending stress and replicate the loading
conditions introduced in the NREL
report~\cite{gage2021laboratory}. Specifically, we compute the neutral
axis of the cross-section and its principal axes. Next, we apply the
bending moment as a traction vector,
$\hat{\vec{t}}=(\hat{t}_{X_1},0,0)^{\top}$, in the $X_1$ direction computed
from the standard flexure formula for unsymmetric bending. The $X_1$
component of the traction vector reads
\begin{equation}
  \hat{t}_{X_1}=\frac{M_{X_2}^{'}X_3^{'}}{I_{X_2}^{'}} - \frac{M_{X_3}^{'}X_2^{'}}{I_{X_3}^{'}},
\end{equation}
where $X_2^{'}$ and $X_3^{'}$ are the principal axes of the cross-section,
$I_{X_2}^{'}$ and $I_{X_3}^{'}$ represent the principle moments of area,
and $M^h=\sqrt{(M_{X_2}^{'})^2+(M_{X_3}^{'})^2}$. The applied moment,
$M^h$, increases linearly from 0.0 to 250.0 kN$\cdot$m over a span of
of 60 seconds to ensure the quasi-static loading condition. In our
simulations, we prevent warping (i.e., deplanation) of the
cross-section at $X_1=0.2$ m (i.e., the cross-section remains
planar). Moreover, we ignore the torque moment on the blade as it is
small, and allow the cross-section at $X_1=0.2$ m to rotate freely around any axis.

\begin{table}[!htb]
\fontsize{9}{11}\selectfont
\begin{center}
\caption{Material properties of the NRT wind turbine blade.}\label{tab:material_prop3}\vspace{2mm}
\begin{tabular}{ p{1.5cm} | p{1.5cm} | p{1.5cm} | p{1.5cm} | p{0.8cm}  } \hline
& ${}^0E$ [MPa] & ${}^0\mu^*$ [MPa] & ${}^0\kappa$ [MPa] & ${}^0\nu$ [-]\\  \hline
Spar cap & $12.0\times 10^3$ &$3.70\times 10^3$ & $8.70\times 10^3$ &0.27\\ \hline
Shear web & $5.43\times 10^3$ &$1.97\times 10^2$& $3.77\times 10^3$ &0.26\\ \hline
Skin & $4.91\times 10^3$ &$1.82\times 10^2$ & $3.41\times 10^3$ &0.26\\ \hline
\end{tabular}
\end{center}
\end{table}
In our constitutive model, we selected [+45/-45] layups of E-glass
fiber-reinforced composite materials. The spar caps are designed with
a thickness of 11 mm, and the shear webs are set at a thickness of
16.7 mm. The thickness of the skin is established at 15.1 mm. These
dimensions are approximated from the actual NRT data. For the shear
webs, we utilize volume-averaged effective properties derived from a
sandwich structure core of end-grain balsa wood as detailed
in~\cite{norris1985differential}. The macroscale material properties
of the E-glass fiber-reinforced composites and sandwich structures
used in our simulations are summarized in
Table~\ref{tab:material_prop3}.

\begin{figure}[!htb]
\centering
{\includegraphics[width=0.8\textwidth]{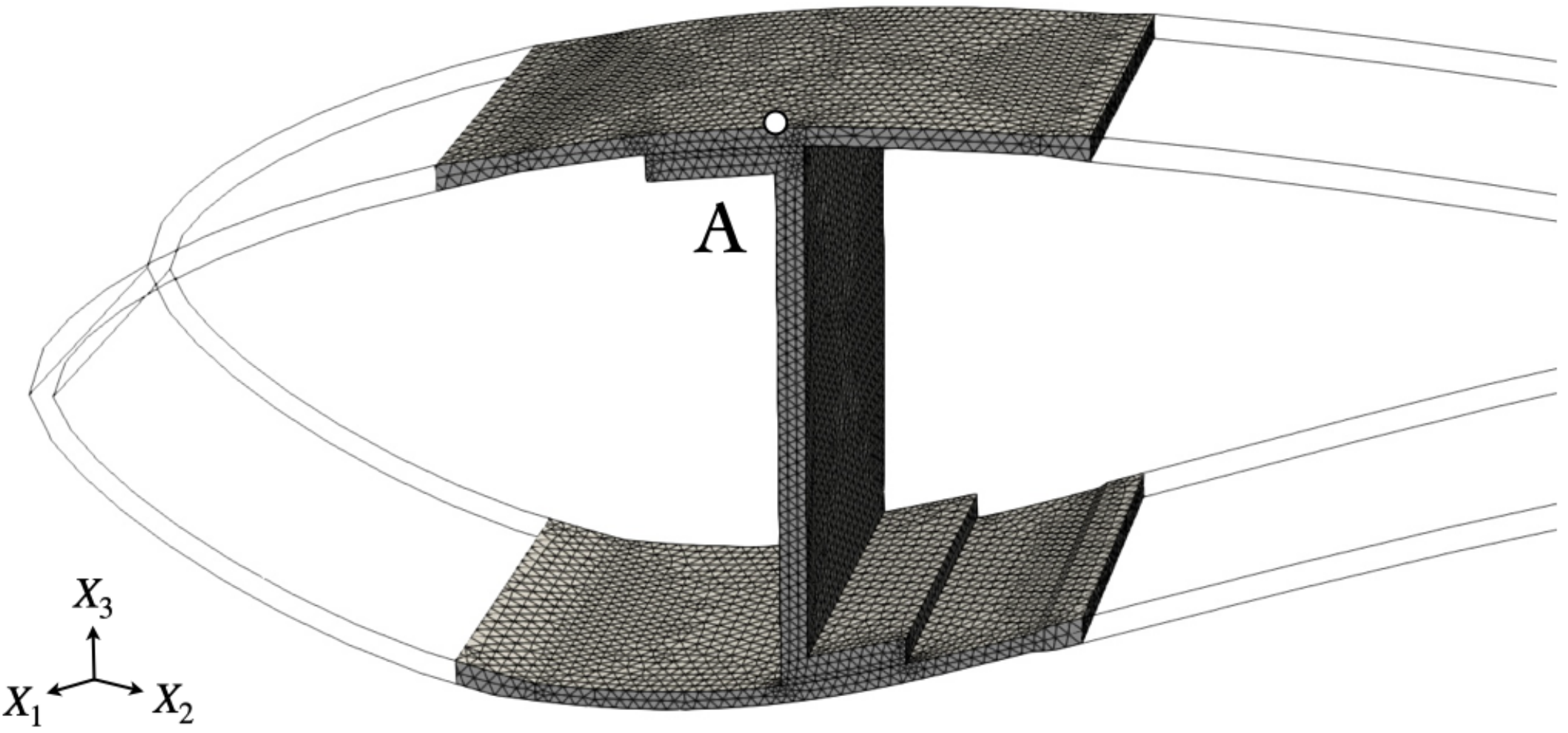}}
\caption{A section of the mesh highlighting the main shear web. Point $A$ marks the center of the spar cap.}\label{fig:ams_par1}
\end{figure}
The macroscale mesh of the NRT blade section consists of 37,667 nodes,
156,600 linear tetrahedral finite elements, 954 cohesive elements, and
109,251 DOFs. This mesh configuration is chosen based on the mesh
refinement study. We select the macroscopic mesh density so as to capture the
main deformation modes. For illustration, Fig.~\ref{fig:ams_par1}
depicts the finite element mesh assigned to the spar caps and the main
shear web of the blade section. The center of the spar cap is
indicated by a white point labeled $A$. 

\begin{figure}[!htb]
\centering{\includegraphics[width=0.49\textwidth]{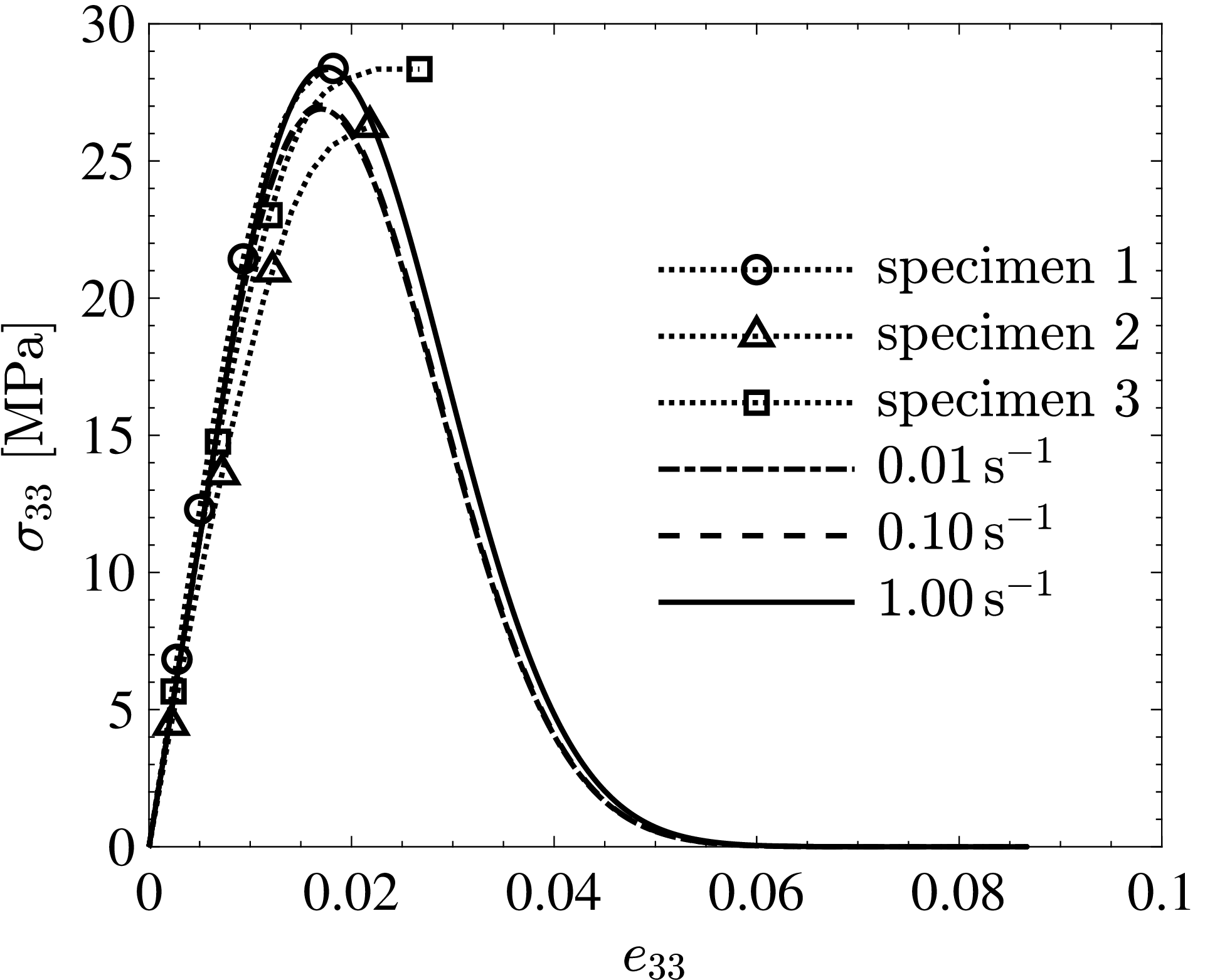}}
\caption{Calibration of the constitutive model with experimental data for three loading rates.}\label{fig::epoxy_rate_sensitivity}
\end{figure}
The top and bottom adhesive layers between the spar caps and the main
shear web are manufactured from a structural adhesive. The shear web
at the trailing edge is assumed to be perfectly bonded to the rest of the
blade. Fig.~\ref{fig::NRT_macrob} shows an example RUC with dimension of
$100\times100\times100\,\mu\text{m}^3$ which consists of a structural epoxy
adhesive containing 4 voids with a diameter of 19.3 $\mu\text{m}$. Volume fraction of the voids is $c_p=1.51\%$ and the voids are randomly
distributed within each RUC using the packing code
$Rocpack$~\cite{stafford2010using}. The diameters and volume fraction
of voids correspond to values of typical manufacturing defects
reported in the literature~\cite{sanchez2014modeling}. The material
properties are calibrated from experimental data for a structural
adhesive epoxy under tensile loading
conditions~\cite{monteiro2015experimental}. Fig.~\ref{fig::epoxy_rate_sensitivity}
shows the normal stress-strain behavior (i.e.,
$\sigma_{33}$--$e_{33}$) of the epoxy under uniaxial tension at
different strain rates. The damage parameters are chosen to ensure
that the peak and failure stress fall within the ranges reported in
the
literature~\cite{monteiro2015experimental,carlberger2009influence}. Table~\ref{tab:material_prop2}
displays material properties calibrated from experimental data.
\begin{table}[!ht]
\fontsize{9}{11}\selectfont
\begin{center}
\caption{Material properties of the epoxy structural adhesive.}\label{tab:material_prop2}\vspace{2mm}
\begin{tabular}{ p{1.5cm} | p{1.5cm} | p{1.5cm} | p{1.5cm} | p{1.5cm} | p{1.5cm} | p{1.5cm} | p{1.5cm}} \hline
$E$ [MPa] &$\mu^*$ [MPa] & $\kappa$ [MPa] & $\nu$ [MPa] & $Y_{in}$ [MPa] & $p_1$ [-] & $p_2$ [-] & $\mu$ [$\text{s}^{-1}$]\\  \hline
$3.0\times 10^3$&$1.1\times 10^3$ & $3.3\times 10^3$ & 0.35 &0.10 & 10.0 & 1.1 & 100.0\\ \hline
\end{tabular}
\end{center}
\end{table}

Moving forward, a total of 954 RUCs, each linked to a corresponding
cohesive element, are allocated to material points on the
adhesives. Of these, 520 RUCs are assigned to the upper adhesive near
the LP skin, and 434 RUCs to the lower adhesive adjacent to the HP
skin. Each RUC includes 1,619 nodes, 9,016 finite elements, and 4,416
DOFs with average element size of $4.8\,\mu\text{m}$. The mesh
density is determined by refinement study employing average element
sizes from $5.0$ to $4.0\,\mu\text{m}$ under different loading
rates of $\|\llbracket{}^0{\dot{\vec{u}}}\,\rrbracket/l_c\|=1.00,\,
0.10, \,\text{and}\, 0.01\, \text{s}^{-1}$. Additionally, we allocate
4 computing cores to an RUC based on a strong scaling study. We
note that if performed as a single-scale FEM simulation, such a
simulation would contain a total of 8,757,864 elements. This would be
a large simulation with complicated contiguous meshing between two
scales. Such simulations are inherently difficult, especially for
highly nonlinear material behavior (e.g., damage).

\begin{figure}[!htb]
\centering
{\includegraphics[width=0.49\textwidth]{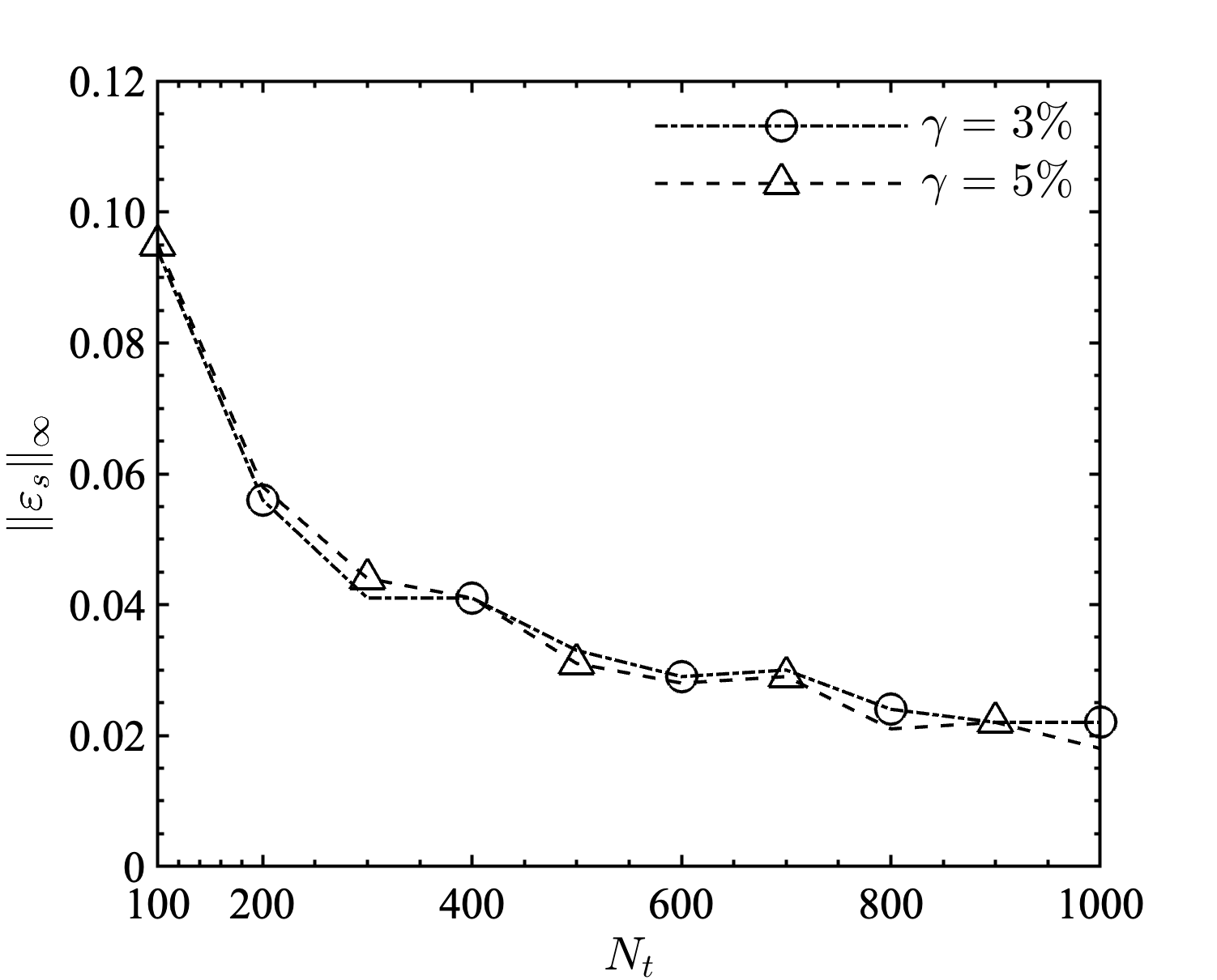}}
\caption{The classification error for nonlinear classifiers trained on epoxy RUC data with varying model tolerances of $\gamma=3\%$ and 5\%.\label{fig:svm_class_epoxy}}
\end{figure}
Next, we generate training samples for structural epoxy (i.e.,
$N_t\in [100,\,1000]$) for each segment, along with $N_c=1000$
quasi-random testing samples (i.e., classification samples) for
each segment. This training sequence is designed to assess the density
of the database. Loading is assigned with a maximum elongation
of $\lambda=0.05 \times l_c$ in Eq.~(\ref{eq221}), based on
the observation that the complete failure of the adhesive occurs near
the 5\% elongation (see Fig.~\ref{fig::epoxy_rate_sensitivity}). The
database, $\mathcal{DB}$, is trained on 1,040 cores. We evaluate the
classification error, Eq.~\eqref{eq:errordatabase}, of $\mathcal{DB}$
for model tolerances $\gamma=3\%$ and $5\%$, to verify that the
density of training samples $N_{t}$ is sufficient. The tighter
modeling tolerance $\gamma$ is important in this example, because the
response of the RUCs is shear-dominated.

\begin{figure}[!htb]
\centering
\subfigure[]{\includegraphics[width=0.35\textwidth]{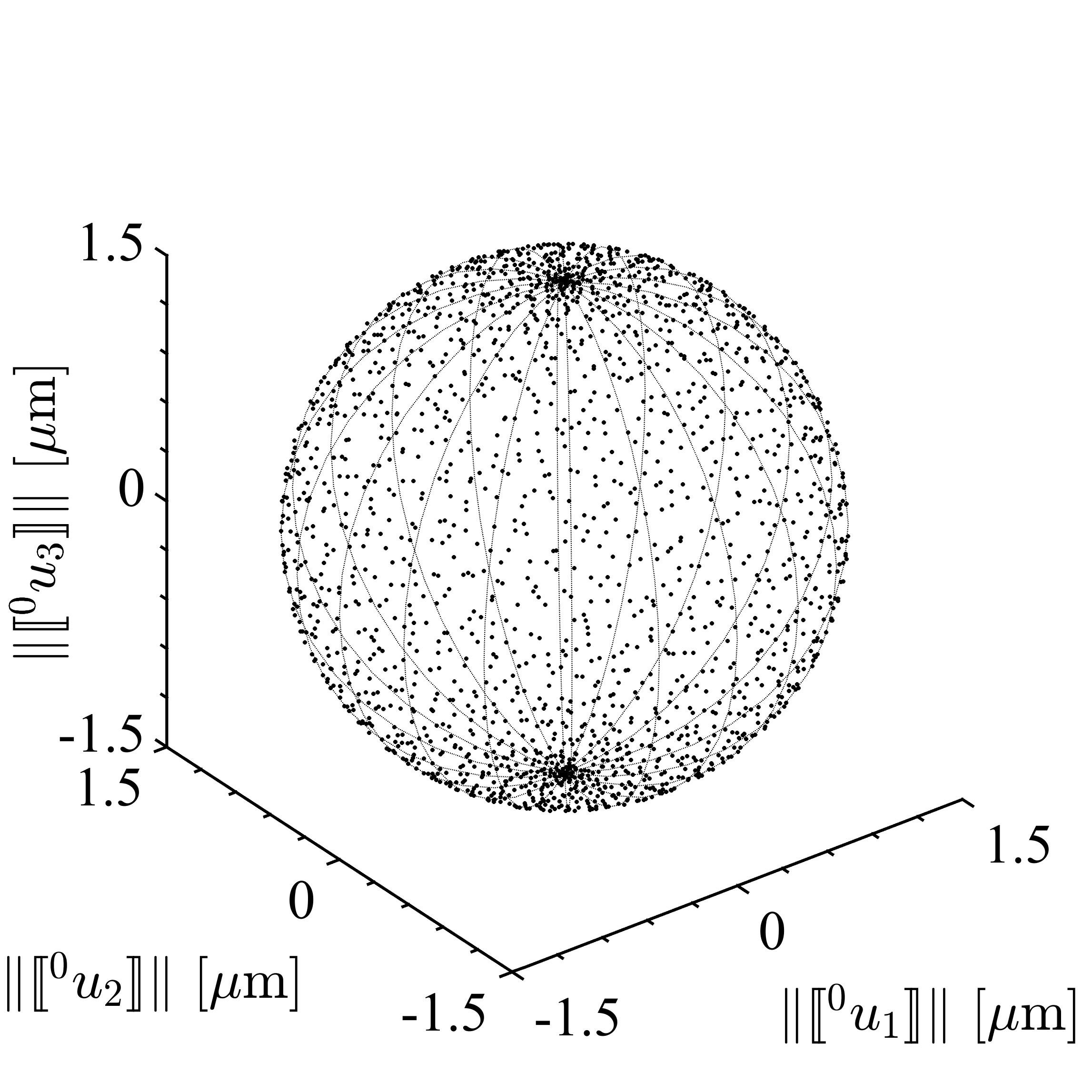}\label{fig:epoxy_svm_errora}}\hspace{2mm}
\subfigure[]{\includegraphics[width=0.60\textwidth]{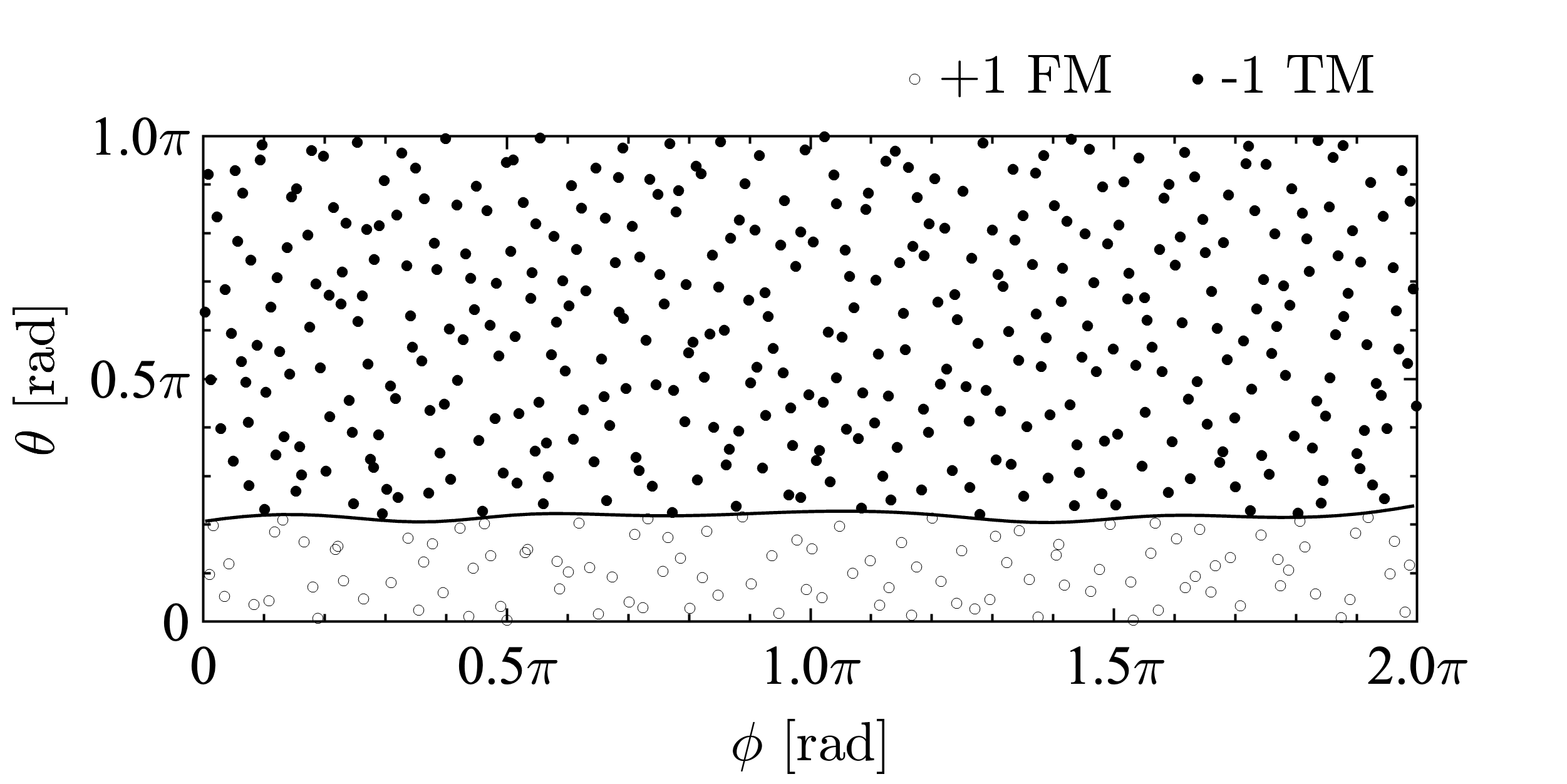}\label{fig:epoxy_svm_errorb}}
\caption{An offline database with $r=30$\% elongation for epoxy RUC with $\gamma=5$\% model tolerance. (a) 3D jump displacement space. (b) 2D spherical coordinate system.\label{fig:epoxy_svm_error}}
\end{figure}
Fig.~\ref{fig:svm_class_epoxy} shows the classification error for both
$\gamma$ values and we select $N_t=500$ for further adaptive multiscale
simulations. In this case we misclassified less than 3.8\%
of samples. Fig.~\ref{fig:epoxy_svm_error} shows the database for
$r=30$\% elongation with $\gamma=5$\% modeling tolerance. As can be seen,
a large portion of the $(\phi,\theta)$ space can be modeled using TM.

\begin{figure}[!htb]
\centering
\subfigure[]
{\includegraphics[width=0.49\textwidth]{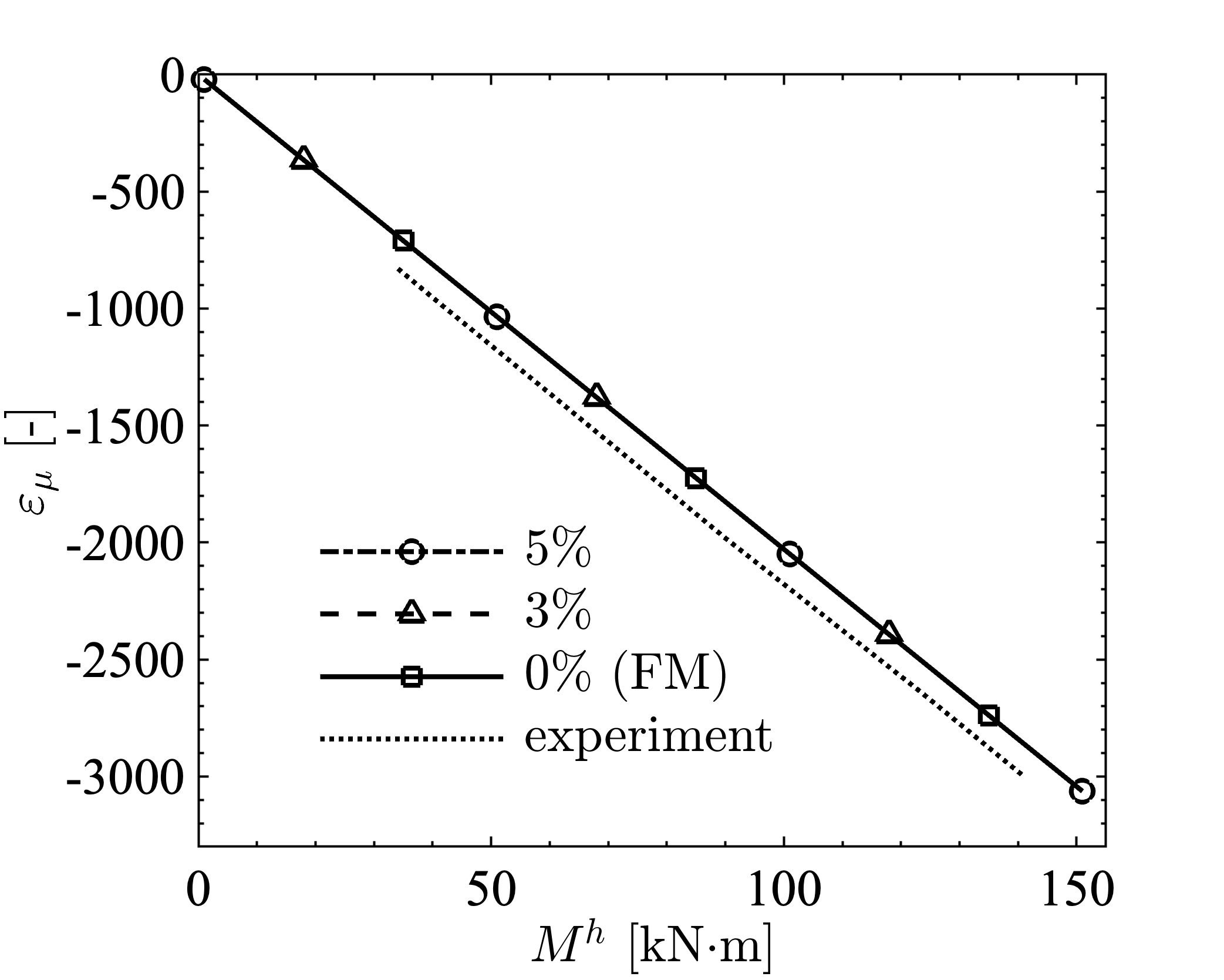}\label{figure:ams_result_1a}}
\subfigure[]
{\includegraphics[width=0.49\textwidth]{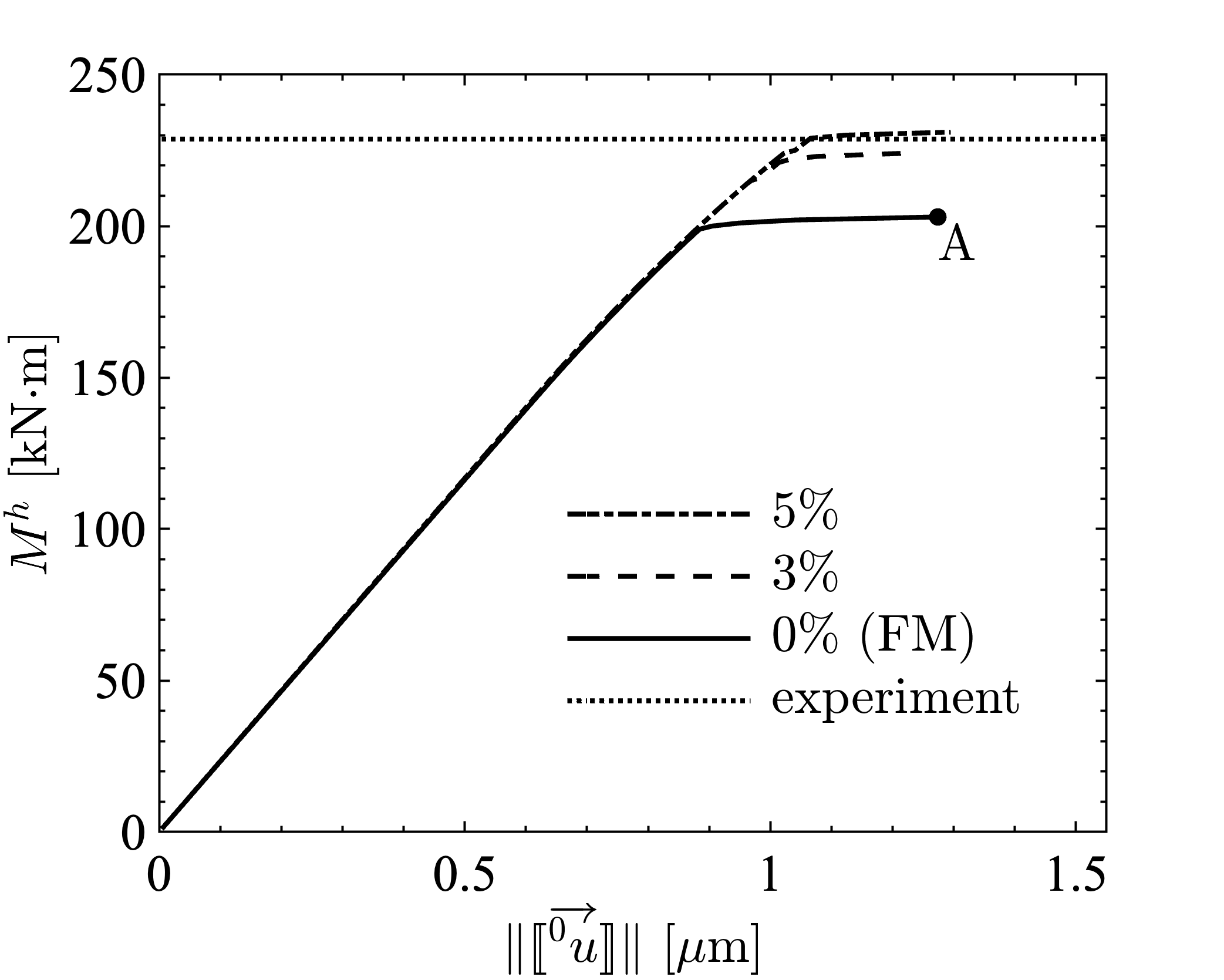}\label{figure:ams_result_1b}}
\caption{Validation of the adaptive multiscale simulations. (a) Bending moment versus the micro-strain. (b) Ultimate bending moment. Note that the micro-strain is computed as $\varepsilon_\mu = \varepsilon_b \times 10^6$.}\label{figure:ams_result_1}
\end{figure}
The multiscale simulations are executed on 1,040 cores, with 1,024
cores dedicated to microscale servers and 16 cores allocated to the
macroscale domain. The simulations are validated against the
experimental measurements of micro-strain and the ultimate bending
moment. The micro-strain $\varepsilon_\mu$ is measured by a strain
gauge mounted on the spar cap at the 2.9 m station. In our simulations,
we compute the bending strain using curvature 
\begin{align}
\varepsilon_b = -\frac{l_{X_3}}{\rho_b},
\end{align}
where $l_{X_3}$ is the distance from the neutral axis and $\rho_b$ is
the local radius of curvature. Taking the bending profile to be
the arc of a circle, we compute the radius of curvature as
$\rho_b=\sfrac{L_s}{\zeta}$, where $\zeta$ is the central angle. The
central angle is estimated in the $X_1-X_3$ plane from the
displacement data around the neutral axis using a small angle
approximation. We use 19 points and compute the central angle as
$\zeta\approx\sfrac{\sqrt{{}^0u_1^2+{}^0u_3^2}}{L_s}$. The maximum value
does not exceed $\rho_b(\max)=$ 0.0038 radians. The micro-strain is
computed at the marked point $A$ in Fig.~\ref{fig:ams_par1} and
compared with the experimental data in
Fig.~\ref{figure:ams_result_1a}. We note the excellent agreement
between the simulations and experiments with micro-strain error in
$\varepsilon_\mu$ of $\approx150$. The micro-strain measurement was
reported up to the applied moment $M^h=$ 140 kN$\cdot$m, which is
within the hyper-elastic response of the blade. Thus, the $\gamma$
tolerance does not the impact the results.

\begin{figure}[!htb]
\centering
\subfigure[]
{\includegraphics[width=0.42\textwidth]{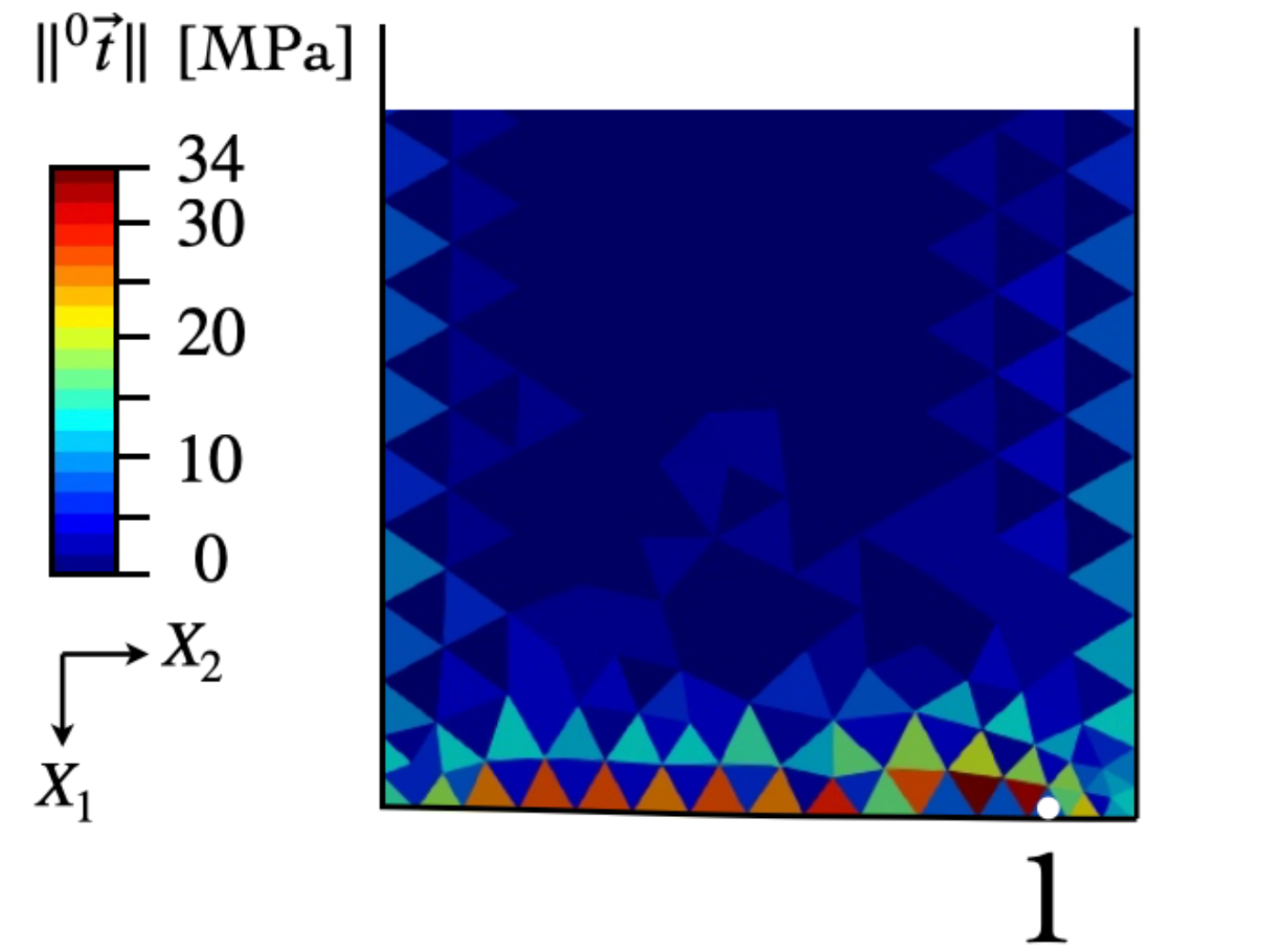}\label{figure:ams_result_3_ruca}} \hspace{5mm}
\subfigure[]
{\includegraphics[width=0.42\textwidth]{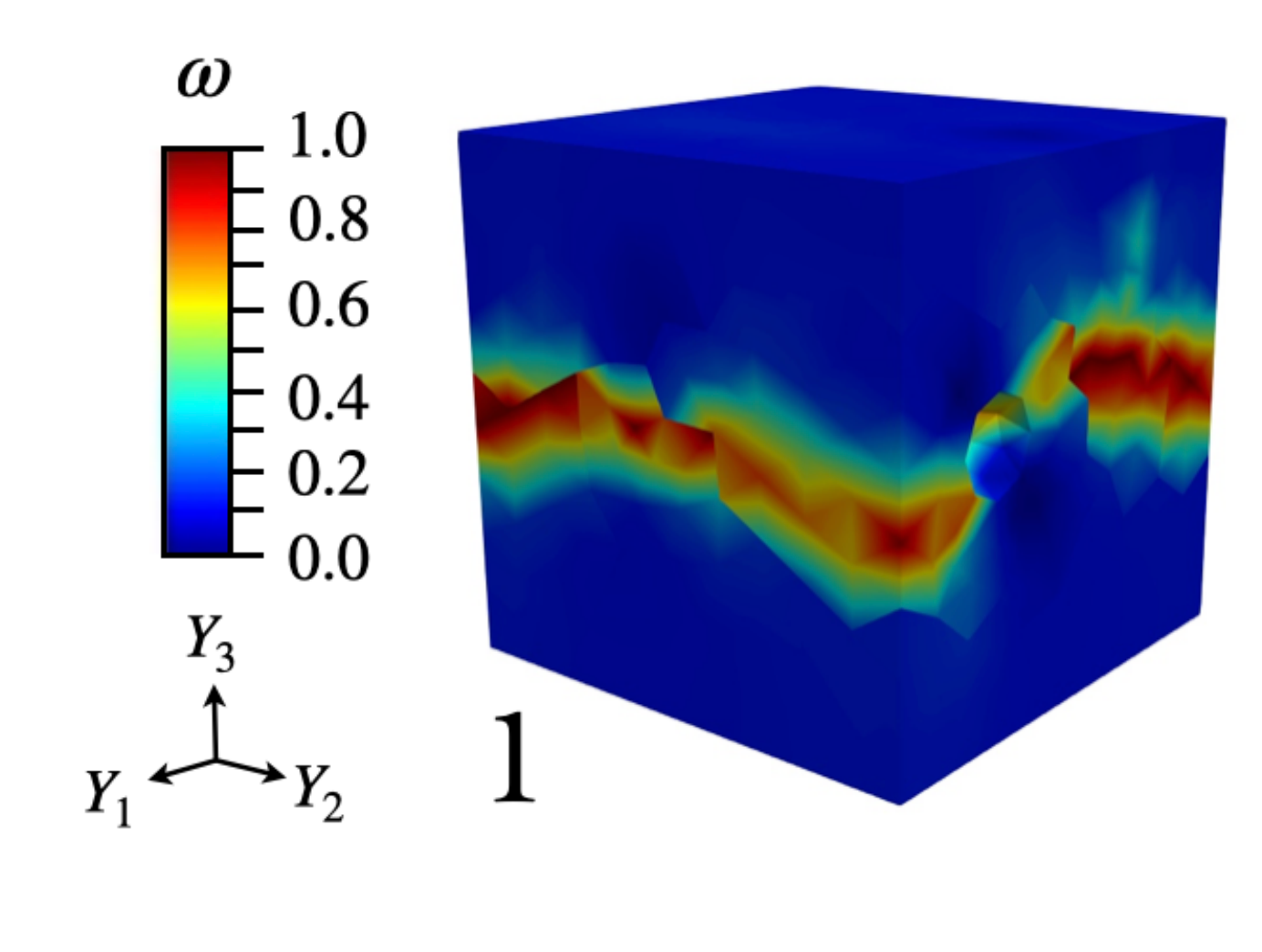}\label{figure:ams_result_3_rucb}}
\caption{Low-pressure side adhesive layer. (a) Magnitude of the traction vector at point A in Fig.~\ref{figure:ams_result_1b}. (b) Localized microscale damage pattern at the marked material point 1 in Fig.~\ref{figure:ams_result_3_ruca}.}\label{figure:ams_result_3_ruc}
\end{figure}
Next, we validate the multiscale simulations against the ultimate
moment. Fig.~\ref{figure:ams_result_1b} shows the applied moment as
a function of the jump displacement. We compute the jump displacement
as an average over the cohesive elements in the first row of the adhesive
layer at the low-pressure interface (see
Fig.~\ref{figure:ams_result_3_ruca} for discretization of the
interface). When the blade failed catastrophically at the 2.9 m
station, the applied bending moment was 228.7 kN$\cdot$m. Our
multiscale simulations with FM predict an ultimate moment of 203
kN$\cdot$m (see Fig.~\ref{figure:ams_result_1b}). That is only 11.2\%
error compared to the experimental measurement. Considering the
complexity of both the experiment and the multiscale simulations, such
a small error is very impressive. Both adaptive multiscale simulations
over-predict the FM results but are closer to the experimental
measurement.

Fig.~\ref{figure:ams_result_3_ruc} displays the norm of the traction
vector and the damage pattern in the failed RUC. We note that the RUCs
fail mostly in shear (see Fig.~\ref{figure:ams_result_3_ruca}). In
our work, the failure of a cohesive element (i.e., RUC) is
characterized by the point where the magnitude of the displacement
jump satisfies $\|\llbracket{}^0\vec{u}\,\rrbracket\|\geq
3\,\mu\text{m}$. This captures nearly the entire material response as
shown in Fig.~\ref{fig::epoxy_rate_sensitivity} (i.e. $e_{33}\geq$ 0.03 for
complete failure). We note that only a small number of cohesive elements
fail, but even this small number is enough to initiate rapid opening
of the adhesive layer without an increase in bending moment as
depicted in Fig.~\ref{figure:ams_result_1b}. For $\gamma$ = 5\%, 3\%,
and 0\% (i.e., FM) models, we have detected 2, 4, and 3 fully failed
cohesive elements, respectively.

\begin{figure}[!htb]
\centering
{\includegraphics[width=0.50\textwidth]{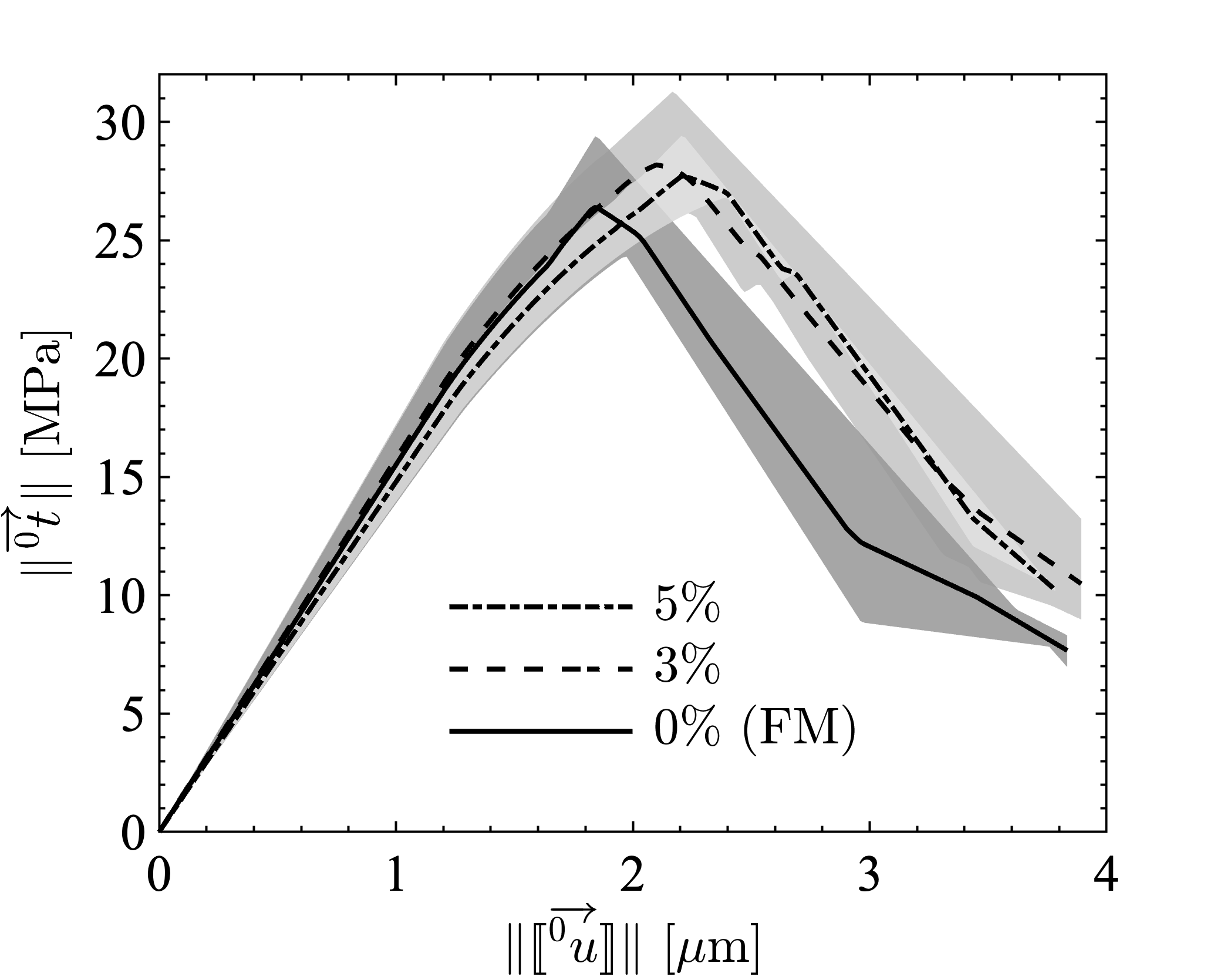}}
\caption{Average traction-separation response of failed cohesive elements based on the modeling tolerance.\label{figure:ams_result_3}}
\end{figure}
Fig.~\ref{figure:ams_result_3} shows the averaged
traction-separation response with the observed spread among elements
represented by different shades of gray around the mean. The gray
regions are represented by min/max values for the individual $\gamma$
simulations. As expected, the maximum averaged traction vector values
are similar to the maximum stress values in
Fig.~\ref{fig::epoxy_rate_sensitivity}.

\begin{table}[!htb]
\fontsize{9}{14}\selectfont
\begin{center}
\caption{The speedup of adaptive multiscale simulations of NRT blade.}\label{tab:application_result_all}\vspace{2mm}
\begin{tabular}{ p{1.0cm} | p{1.6cm} | p{1.6cm} } \hline
$\gamma$ [\%]  &$N_T/N_{COH}$  &  $T_{FM}/T_{AM}$ \\  \hline
5 & 0.9937 & 2.75 \\ \hline
3 & 0.9895 & 2.45 \\ \hline
\end{tabular}
\end{center}
\end{table}
Next, we present the acceleration data for our adaptive
simulations. Table~\ref{tab:application_result_all} shows speedup and
the ratio of TM versus FM. We can see that the speedup is quite large
compared to that of the DCB example presented in
Section~\ref{DCB_section}. We also note that introducing a
third model (e.g., a phenomenological cohesive law) would further
accelerate the multiscale simulations.

\begin{figure}[!htb]
\centering
\subfigure[]{\includegraphics[width=0.85\textwidth]{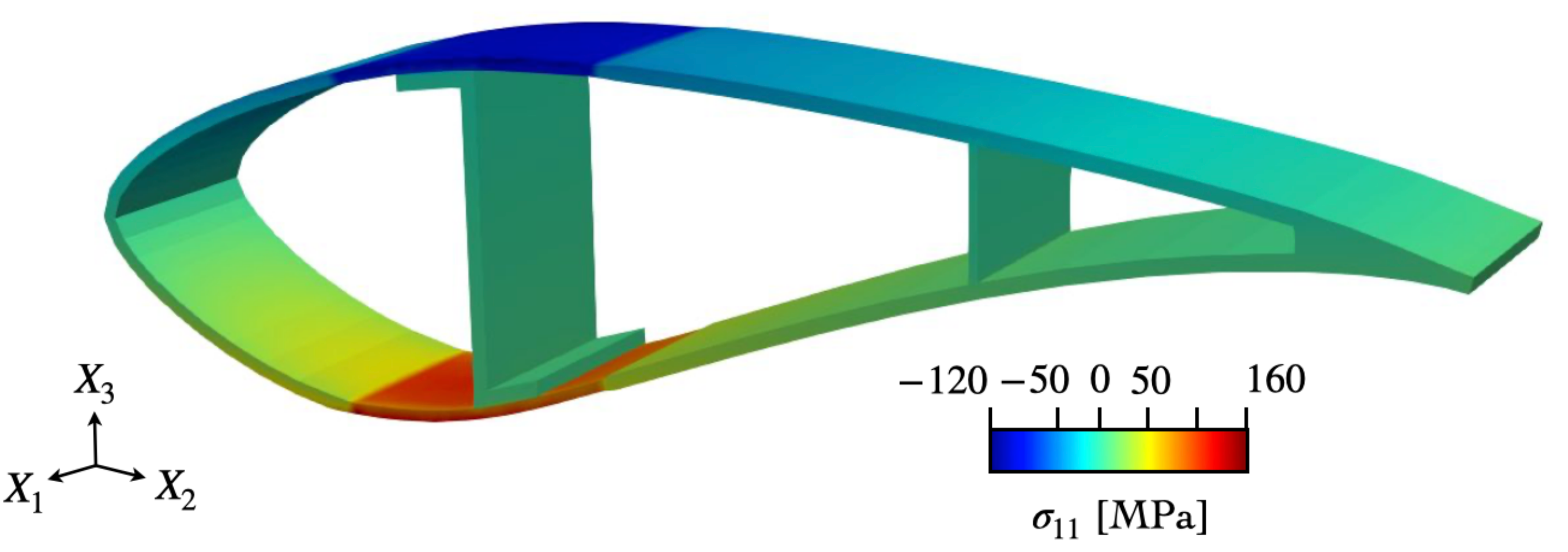}\label{figure:ams_result_3_bladea}}
\subfigure[]{\includegraphics[width=0.85\textwidth]{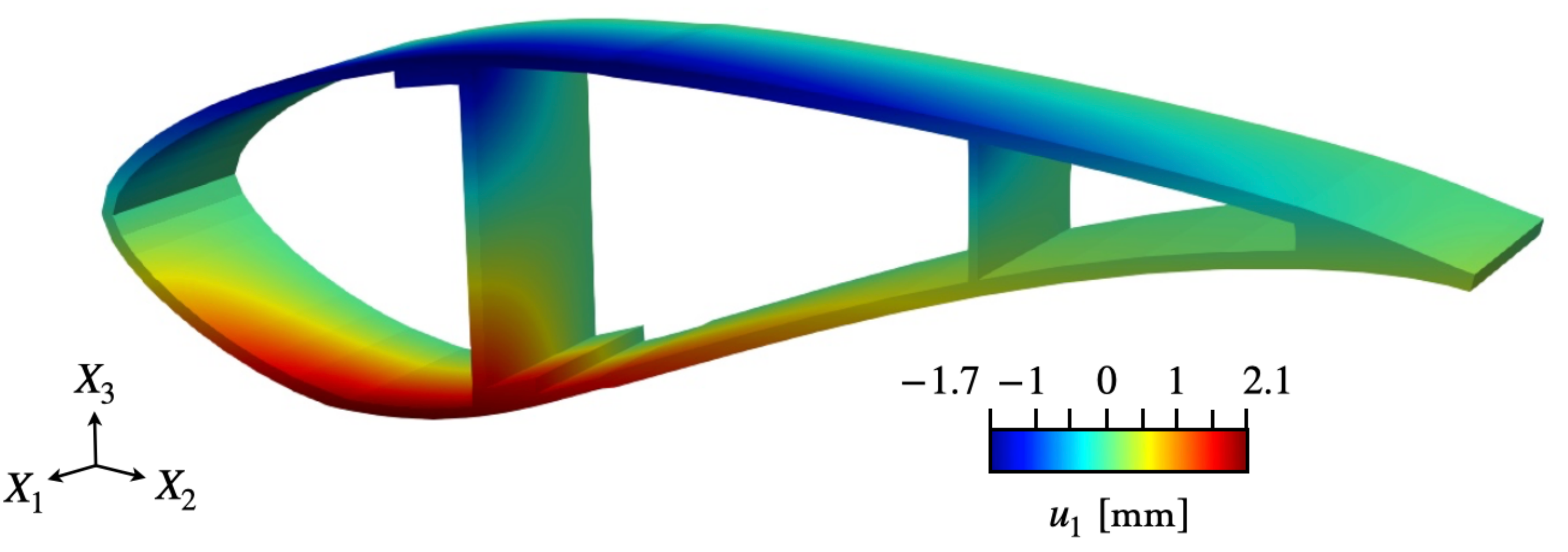}\label{figure:ams_result_3_bladeb}}
\caption{Macroscopic response of the NRT blade. (a) Stress component in the $X_1$ direction. (b) Displacement component in the $X_1$ direction.\label{figure:ams_result_3_blade}}
\end{figure}
We conclude this example with the macroscopic response of the NRT
blade at the end of the loading history (see marked point A in
Fig.~\ref{figure:ams_result_1b}). Fig.~\ref{figure:ams_result_3_blade}
displays the $\sigma_{11}$ component of the Cauchy stress tensor and
the $u_1$ component of the displacement vector, respectively. Tensile stress of $\approx 160$ MPa and compressive stress of $\approx
120$ MPa are observed at the high-pressure and low-pressure sides of
the blade. As expected, the most stress is carried by the spar caps
above the main shear web. The skin also carries some load, but the
trailing edge is mostly stress-free. The blade deformation is quite
nonuniform, with the leading and trailing edges deforming slightly as
they are both close to the neutral axis. We note again that the shear
web at the trailing edge is perfectly bonded to the rest of the
structure.
\section{Conclusion}\label{sec:7}
We present a new adaptive and parallel multiscale solver for
large-scale engineering applications, leveraging support vector
regression and a new networking library. This adaptive approach
utilizes computational homogenization of interfaces and two microscale
models based on an offline database to predict highly nonlinear
material behavior across scales, thereby reducing computational time
while ensuring high-fidelity results. Our networking library
facilitates the dynamic selection of microscale models within a
high-performance computing setting, significantly enhancing simulation
feasibility. The hierarchically parallel multiscale solver is
carefully verified and subsequently validated for a large
engineering problem. In particular, we simulate a complex real-world
engineering application involving the ultimate failure of a large
wind turbine blade. This work represents an advancement in multiscale
simulations, offering a comprehensive solution to computational
multiscale demands both through model selection and high-performance
computing. Finally, we highlight the potential for introducing a whole
library of models (i.e., phenomenological, ROM, etc.) that could be
considered in future studies to improve computational performance
while retaining the predictability of industrially relevant large
multiscale simulations.  
\section*{Acknowledgment}
This work was supported by the Department of Energy, National Nuclear
Security Administration, under the award No. DENA0002377 as part of
the Predictive Science Academic Alliance Program II.

\end{document}